\PassOptionsToPackage{unicode}{hyperref}
\PassOptionsToPackage{hyphens}{url}
\PassOptionsToPackage{dvipsnames,svgnames,x11names}{xcolor}
\documentclass[
  10pt,
  letterpaper,
]{article}
\usepackage{xcolor}
\usepackage{amsmath,amssymb}
\usepackage{iftex}
\ifPDFTeX
  \usepackage[T1]{fontenc}
  \usepackage[utf8]{inputenc}
  \usepackage{textcomp} % provide euro and other symbols
\else % if luatex or xetex
  \usepackage{unicode-math} % this also loads fontspec
  \defaultfontfeatures{Scale=MatchLowercase}
  \defaultfontfeatures[\rmfamily]{Ligatures=TeX,Scale=1}
\fi
\usepackage{lmodern}
\ifPDFTeX\else
\fi
\IfFileExists{upquote.sty}{\usepackage{upquote}}{}
\IfFileExists{microtype.sty}{% use microtype if available
  \usepackage[]{microtype}
  \UseMicrotypeSet[protrusion]{basicmath} % disable protrusion for tt fonts
}{}
\makeatletter
\@ifundefined{KOMAClassName}{% if non-KOMA class
  \IfFileExists{parskip.sty}{%
    \usepackage{parskip}
  }{% else
    \setlength{\parindent}{0pt}
    \setlength{\parskip}{6pt plus 2pt minus 1pt}}
}{% if KOMA class
  \KOMAoptions{parskip=half}}
\makeatother
\makeatletter
\ifx\paragraph\undefined\else
  \let\oldparagraph\paragraph
  \renewcommand{\paragraph}{
    \@ifstar
      \xxxParagraphStar
      \xxxParagraphNoStar
  }
  \newcommand{\xxxParagraphStar}[1]{\oldparagraph*{#1}\mbox{}}
  \newcommand{\xxxParagraphNoStar}[1]{\oldparagraph{#1}\mbox{}}
\fi
\ifx\subparagraph\undefined\else
  \let\oldsubparagraph\subparagraph
  \renewcommand{\subparagraph}{
    \@ifstar
      \xxxSubParagraphStar
      \xxxSubParagraphNoStar
  }
  \newcommand{\xxxSubParagraphStar}[1]{\oldsubparagraph*{#1}\mbox{}}
  \newcommand{\xxxSubParagraphNoStar}[1]{\oldsubparagraph{#1}\mbox{}}
\fi
\makeatother

\usepackage{longtable,booktabs,array}
\usepackage{calc} % for calculating minipage widths#
\usepackage{etoolbox}
\makeatletter
\patchcmd\longtable{\par}{\if@noskipsec\mbox{}\fi\par}{}{}
\makeatother
\IfFileExists{footnotehyper.sty}{\usepackage{footnotehyper}}{\usepackage{footnote}}
\makesavenoteenv{longtable}
\usepackage{graphicx}
\makeatletter
\newsavebox\pandoc@box
\newcommand*\pandocbounded[1]{% scales image to fit in text height/width
  \sbox\pandoc@box{#1}%
  \Gscale@div\@tempa{\textheight}{\dimexpr\ht\pandoc@box+\dp\pandoc@box\relax}%
  \Gscale@div\@tempb{\linewidth}{\wd\pandoc@box}%
  \ifdim\@tempb\p@<\@tempa\p@\let\@tempa\@tempb\fi% select the smaller of both
  \ifdim\@tempa\p@<\p@\scalebox{\@tempa}{\usebox\pandoc@box}%
  \else\usebox{\pandoc@box}%
  \fi%
}
\def\fps@figure{htbp}
\makeatother

\NewDocumentCommand\citeproctext{}{}
\NewDocumentCommand\citeproc{mm}{%
  \begingroup\def\citeproctext{#2}\cite{#1}\endgroup}
\makeatletter
 \let\@cite@ofmt\@firstofone
 \def\@biblabel#1{}
 \def\@cite#1#2{{#1\if@tempswa , #2\fi}}
\makeatother
\newlength{\cslhangindent}
\newlength{\csllabelwidth}
\newenvironment{CSLReferences}[2] % #1 hanging-indent, #2 entry-spacing
 {\begin{list}{}{%
  \setlength{\itemindent}{0pt}
  \setlength{\leftmargin}{0pt}
  \setlength{\parsep}{0pt}
  \ifodd #1
   \setlength{\leftmargin}{\cslhangindent}
   \setlength{\itemindent}{-1\cslhangindent}
  \fi
  \setlength{\itemsep}{#2\baselineskip}}}
 {\end{list}}
\usepackage{calc}

\newcommand{\CSLLeftMargin}[1]{\parbox[t]{\csllabelwidth}{\strut#1\strut}}
\newcommand{\CSLRightInline}[1]{\parbox[t]{\linewidth - \csllabelwidth}{\strut#1\strut}}

\usepackage{PRIMEarxiv}
\usepackage{algorithm}
\usepackage{algpseudocode}
\usepackage{amsmath}
\usepackage{amssymb}
\usepackage{booktabs}
\usepackage{float}
\usepackage{hyperref}
\usepackage{url}
\usepackage{microtype}
\usepackage[export]{adjustbox}
\usepackage{needspace}
\BeforeBeginEnvironment{longtable}{\needspace{0.4\textheight}}
\let\origincludegraphics\includegraphics
\renewcommand{\includegraphics}[2][]{\origincludegraphics[max height=0.85\textheight,keepaspectratio,#1]{#2}}
\makeatletter
\AtBeginDocument{%
  \def\@author{%
    Mohammad Talebi-Kalaleh \\
    {\mdseries\small Department of Civil and Environmental Engineering} \\
    {\mdseries\small University of Alberta, Edmonton, Canada}
    \And
    Qipei Mei\thanks{Corresponding author: \texttt{qipei.mei@ualberta.ca}} \\
    {\mdseries\small Department of Civil and Environmental Engineering} \\
    {\mdseries\small University of Alberta, Edmonton, Canada} \\
  }%
}
\makeatother
\makeatletter
\@ifpackageloaded{caption}{}{\usepackage{caption}}
\AtBeginDocument{%
\ifdefined\contentsname
  \renewcommand*\contentsname{Table of contents}
\else
  \newcommand\contentsname{Table of contents}
\fi
\ifdefined\listfigurename
  \renewcommand*\listfigurename{List of Figures}
\else
  \newcommand\listfigurename{List of Figures}
\fi
\ifdefined\listtablename
  \renewcommand*\listtablename{List of Tables}
\else
  \newcommand\listtablename{List of Tables}
\fi
\ifdefined\figurename
  \renewcommand*\figurename{Figure}
\else
  \newcommand\figurename{Figure}
\fi
\ifdefined\tablename
  \renewcommand*\tablename{Table}
\else
  \newcommand\tablename{Table}
\fi
}
\@ifpackageloaded{float}{}{\usepackage{float}}
\floatstyle{ruled}
\@ifundefined{c@chapter}{\newfloat{codelisting}{h}{lop}}{\newfloat{codelisting}{h}{lop}[chapter]}
\floatname{codelisting}{Listing}

\makeatother
\makeatletter
\@ifpackageloaded{caption}{}{\usepackage{caption}}
\@ifpackageloaded{subcaption}{}{\usepackage{subcaption}}
\makeatother
\usepackage{bookmark}
\IfFileExists{xurl.sty}{\usepackage{xurl}}{} % add URL line breaks if available
\hypersetup{
  pdftitle={An Open-Source Framework for Coupled Vehicle-Bridge Interaction Analysis Using OpenSees},
  pdfauthor={Mohammad Talebi-Kalaleh; Qipei Mei},
  pdfkeywords={vehicle-bridge interaction, OpenSees, iterative
partitioned coupling, moving mass, moving load, indirect bridge
monitoring, open-source software},
  colorlinks=true,
  linkcolor={blue},
  filecolor={Maroon},
  citecolor={Blue},
  urlcolor={Blue},
  pdfcreator={LaTeX via pandoc}}

\title{An Open-Source Framework for Coupled Vehicle-Bridge Interaction
Analysis Using OpenSees}
\author{Mohammad Talebi-Kalaleh \and Qipei Mei}
\date{2026-02-16}
\begin{document}
\maketitle
\begin{abstract}
Vehicle-bridge interaction (VBI) is important for simulating bridge
response under moving vehicular loads and supports applications such
as dynamic amplification studies, weigh-in-motion, and indirect bridge
monitoring. Although VBI theory is well established, many existing implementations use custom finite
element code or research-specific solvers, which limits their reuse.
This paper presents an open-source Python framework for VBI analysis
built on OpenSees. The bridge and vehicle are modeled as separate
OpenSees subsystems and connected through an iterative scheme that
exchanges displacement and force values at each time step until
convergence. Five vehicle model types are supported, from a single-axle
spring-mass system to two-axle composite half-cars with body pitch and
separate tyre and suspension elements. A decoupled mode is also
provided: the vehicle static weight is applied as a moving load on the
bridge, and the resulting bridge motion is then used as base excitation
for the vehicle. Validation against three published benchmarks
(quarter-car, half-car with pitch, and full composite models) shows
close agreement, with $R^{2}$ above 0.998 in all cases. A parametric
study reports the accuracy of the decoupled mode as a function of
vehicle-to-bridge mass ratio, span length, speed, road roughness class,
and background traffic density, and indicates when the decoupled mode
is adequate and when full coupling is needed. The complete framework
and benchmark configurations are released as open-source software to
support reproducible research in vehicle-bridge dynamics.
\end{abstract}


\section{Introduction}\label{sec-introduction}

The dynamic interaction between moving vehicles and bridge structures is
a classical problem in structural engineering that has attracted
sustained research attention for more than half a century
{[}\citeproc{ref-Fryba1999}{1},\citeproc{ref-Yang2004book}{2}{]}. When a
vehicle traverses a bridge, the coupled dynamics of the two subsystems
produce responses that differ, sometimes substantially, from those
predicted by simpler models that treat the vehicle as a static or moving
force. The vehicle-bridge interaction (VBI) problem supports a range of
engineering applications, including: (i) evaluating the dynamic
amplification of bridge responses for design and assessment, (ii) bridge
weigh-in-motion systems that infer vehicle axle loads from bridge
response measurements, and (iii) extracting bridge condition information
from the dynamic response of instrumented vehicles passing over the
bridge, commonly referred to as indirect or drive-by bridge monitoring
{[}\citeproc{ref-Yang2004indirect}{3},\citeproc{ref-Malekjafarian2015review}{4}{]}.

The theoretical treatment of VBI dates back to the early works of Fryba
{[}\citeproc{ref-Fryba1999}{1}{]}, who provided analytical solutions for
beams subjected to moving forces and moving masses. Yang, Yau, and Wu
{[}\citeproc{ref-Yang2004book}{2}{]} later developed a comprehensive VBI
element formulation in which the vehicle degrees of freedom are
condensed into the bridge element matrices at the contact points,
enabling a monolithic solution of the coupled system. This formulation
was extended to asymmetric two-axle (half-car) vehicles with body pitch
by Yang et al. {[}\citeproc{ref-Yang2019}{5}{]}, and has served as a
primary reference for validating VBI implementations. Parallel efforts
by Green and Cebon {[}\citeproc{ref-Green1994}{6}{]}, Zhu and Law
{[}\citeproc{ref-Zhu2002}{7}{]}, and others established iterative
partitioned coupling as an alternative to the monolithic approach. In
the partitioned scheme, the bridge and vehicle are solved as independent
subsystems with interaction forces and displacements exchanged
iteratively at each time step until convergence is achieved. This
approach offers greater flexibility, as the subsystems can use different
solvers, meshes, and integration parameters.

These applications span the full spectrum of VBI use, from forward
simulation for design and dynamic amplification studies to inverse
analysis for weigh-in-motion and condition assessment. In particular,
the growth of drive-by monitoring has recently expanded research
interest in VBI simulation. Yang et al.
{[}\citeproc{ref-Yang2004indirect}{3}{]} first demonstrated that bridge
natural frequencies appear in the spectrum of a passing vehicle's
acceleration response, laying the theoretical foundation for drive-by
bridge inspection. Subsequent work by OBrien and colleagues
{[}\citeproc{ref-OBrien2014}{8},\citeproc{ref-OBrien2015}{9},\citeproc{ref-Gonzalez2012}{10}{]},
Malekjafarian et al.
{[}\citeproc{ref-Malekjafarian2015review}{4},\citeproc{ref-Malekjafarian2017}{11}{]},
Eshkevari et al. {[}\citeproc{ref-Eshkevari2020}{12}{]}, and Mei et al.
{[}\citeproc{ref-Mei2019}{13},\citeproc{ref-Mei2021}{14}{]} has advanced
the field considerably, developing methods for frequency extraction,
damping identification, mode shape estimation, and damage detection from
drive-by measurements. All of these applications depend critically on
the fidelity of the VBI model used for their development and validation.
The present work targets general VBI analysis rather than any single
application, providing a reusable simulation platform that can serve
design, weigh-in-motion, and monitoring studies alike.

Despite the maturity of VBI theory, its implementation remains a barrier
for many researchers and engineers. Most existing implementations are
coded from scratch using custom finite element routines or analytical
modal superposition, requiring substantial expertise in computational
mechanics. Purpose-built VBI codes are typically maintained within
individual research groups and are rarely shared as reusable,
well-documented software. Meanwhile, established general-purpose finite
element packages such as OpenSees {[}\citeproc{ref-McKenna2011}{15}{]},
which are widely used in the structural engineering community, do not
natively support moving load or moving mass analysis.

This paper addresses this gap by presenting an open-source Python
framework that performs VBI analysis using OpenSees as the underlying
finite element engine. The key idea is to model the bridge and vehicle
as separate OpenSees finite element models, each defined using standard
OpenSees elements (beam-column elements for the bridge, zero-length
spring-dashpot elements for the vehicle), and to couple them externally
through a Python-level iterative algorithm. This approach requires no
modification to the OpenSees source code and no custom element
formulations. At each time step, the framework constructs the vehicle
input (bridge deformation plus road roughness at the axle locations),
solves the vehicle model for one step to obtain updated reaction forces,
maps these forces back onto bridge nodes, solves the bridge model, and
iterates until the bridge response converges. The framework supports
four vehicle model types of increasing complexity, multi-span bridge
configurations, ISO 8608 road roughness generation, and both fully
coupled (iterative) and decoupled (moving load) analysis modes.

The remainder of this paper is organised as follows.
Section~\ref{sec-literature} reviews the theoretical background and
existing methods for VBI analysis. Section~\ref{sec-methodology}
describes the proposed framework in detail, including the bridge and
vehicle finite element formulations, the iterative coupling algorithm,
the road roughness model, and the simplified decoupled approach.
Section~\ref{sec-validation} validates the framework against established
benchmark configurations from the literature.
Section~\ref{sec-parametric} presents a parametric study comparing the
coupled and decoupled approaches across a range of problem parameters.
Section~\ref{sec-discussion} discusses the main observations,
implications, and limitations. Section~\ref{sec-conclusions} summarises
the findings and outlines directions for future development.

\section{Literature Review}\label{sec-literature}

\subsection{Analytical Foundations}\label{analytical-foundations}

The simplest dynamic loading model treats the vehicle as a constant
force moving at speed \(v\), for which Fryba
{[}\citeproc{ref-Fryba1999}{1}{]} derived closed-form solutions using
modal superposition. Because this model ignores the vehicle's inertia,
it cannot capture the feedback between the vehicle's vibration and the
bridge's deformation, which is the defining feature of the
vehicle-bridge interaction (VBI) problem.

\begin{figure}[htbp]

\centering{

\includegraphics[width=0.9\linewidth,height=\textheight,keepaspectratio]{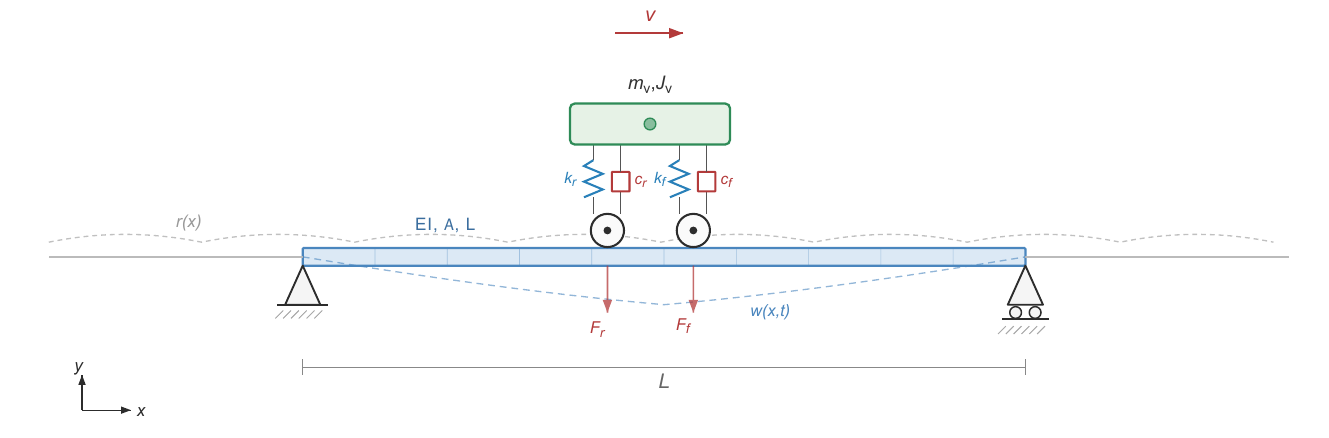}

}

\caption{\label{fig-vbi-schematic}Schematic of the vehicle-bridge
interaction problem. A multi-DOF vehicle traverses an Euler-Bernoulli
beam bridge at velocity \(v\), with interaction forces \(F_r\) and
\(F_f\) transmitted at the rear and front axle positions through shape
function interpolation.}

\end{figure}

In the VBI formulation, two physical mechanisms act simultaneously.
First, the vehicle undergoes \emph{base excitation}: each axle follows
the bridge surface, whose vertical profile equals the sum of the
structural deflection \(w(x,t)\) and the road surface roughness
\(r(x)\). The resulting contact-point displacement

\begin{equation}\phantomsection\label{eq-contact-disp}{
u_c(t) = w\!\bigl(x_a(t),\, t\bigr) + r\!\bigl(x_a(t)\bigr)
}\end{equation}

where \(x_a(t)\) is the axle position, drives the vehicle's equations of
motion from below. Second, the bridge experiences the vehicle as a
\emph{moving dynamic load}: the spring, dashpot, and inertial forces
that the vehicle transmits at each contact point act as time-varying
forces on the beam. Because the bridge deflection depends on the vehicle
forces, which in turn depend on the bridge deflection, the two
subsystems must be solved together at each time step.
Figure~\ref{fig-vbi-schematic} illustrates this coupled problem for a
two-axle vehicle traversing a simply supported beam.

\subsection{Vehicle Models}\label{vehicle-models}

\begin{figure}[htbp]

\centering{

\includegraphics[width=0.95\linewidth,height=\textheight,keepaspectratio]{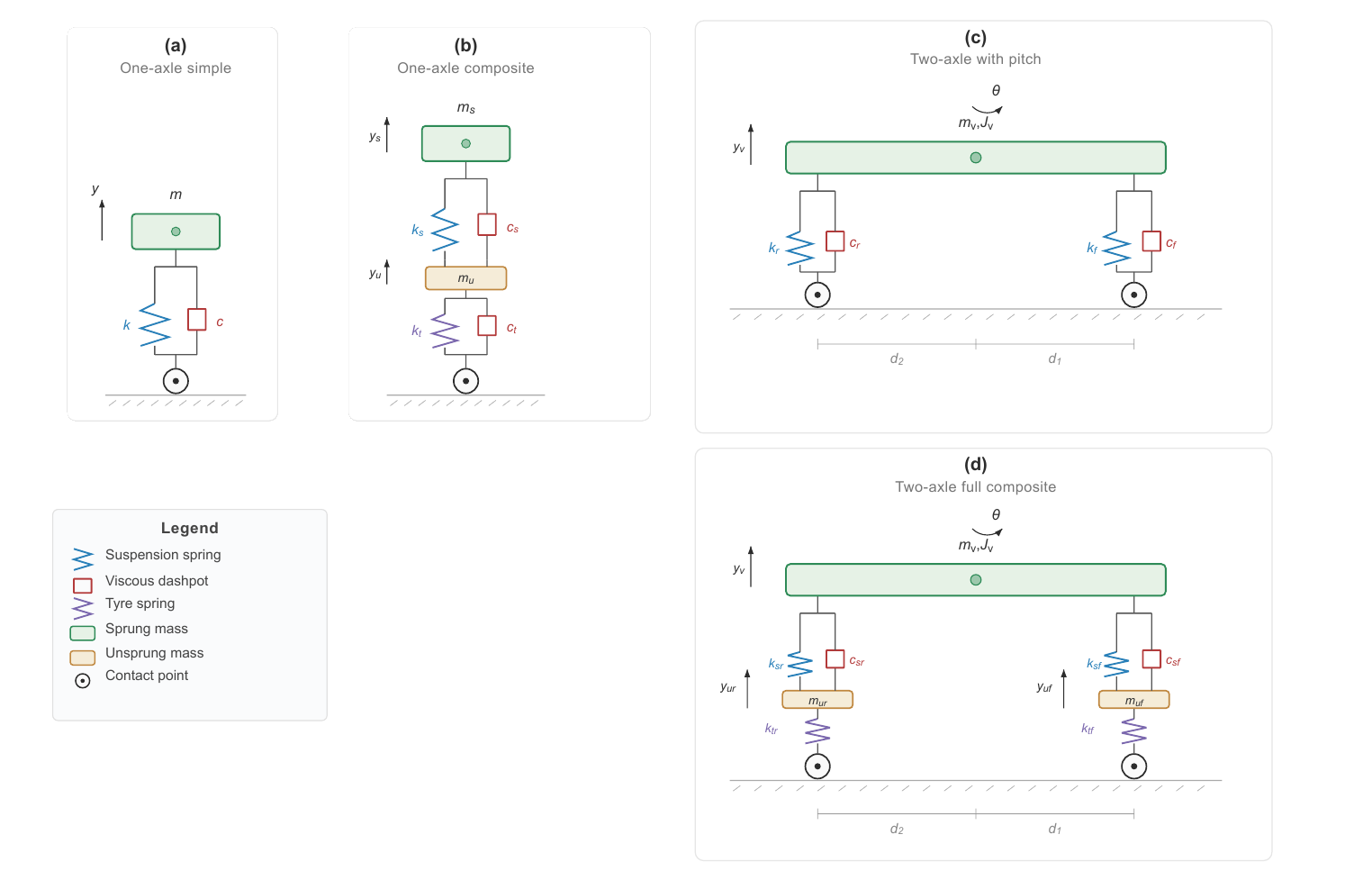}

}

\caption{\label{fig-vehicle-models}Vehicle models supported by the
framework: (a) single-axle simple, (b) quarter-car composite, (c)
half-car with pitch, and (d) full composite half-car with axle masses.}

\end{figure}

The framework supports four vehicle model types of increasing
complexity, illustrated in Figure~\ref{fig-vehicle-models}. All models
are driven by the contact-point displacement \(u_c\) defined in
Equation~\ref{eq-contact-disp}, and all transmit a contact force back to
the bridge that equals the reaction at the constrained axle node. The
governing equations for each model are given below; gravity loads
produce a static component that is computed once at the start of the
analysis and omitted from the dynamic equations that follow.

The \textbf{single-axle simple model} (Figure~\ref{fig-vehicle-models}
a) represents the vehicle as a lumped mass \(m\) connected to the
contact point through a spring \(k\) and dashpot \(c\). With the mass
displacement denoted \(y\), the equation of motion is

\begin{equation}\phantomsection\label{eq-simple-eom}{
m\,\ddot{y} + c\,\dot{y} + k\,y = c\,\dot{u}_c + k\,u_c
}\end{equation}

and the contact force transmitted to the bridge is
\(F_c = k(y - u_c) + c(\dot{y} - \dot{u}_c)\). Yang and Yau
{[}\citeproc{ref-Yang2004book}{2}{]} used this model to derive
analytical VBI solutions and to validate finite element formulations.

The \textbf{quarter-car model} (Figure~\ref{fig-vehicle-models} b)
separates the vehicle into a sprung mass \(m_s\) (body) and an unsprung
mass \(m_u\) (axle/wheel assembly), connected through a suspension
(\(k_s\), \(c_s\)). The axle is connected to the contact point through a
tyre (\(k_t\), \(c_t\)). This two-degree-of-freedom model captures both
the low-frequency body bounce and the higher-frequency axle hop, making
it the standard model for drive-by bridge monitoring studies
{[}\citeproc{ref-OBrien2015}{9},\citeproc{ref-Eshkevari2020}{12}{]}. The
equations of motion are

\begin{equation}\phantomsection\label{eq-quartercar-eom}{
\begin{bmatrix} m_u & 0 \\ 0 & m_s \end{bmatrix}
\begin{Bmatrix} \ddot{y}_u \\ \ddot{y}_s \end{Bmatrix}
+\begin{bmatrix} c_t + c_s & -c_s \\ -c_s & c_s \end{bmatrix}
\begin{Bmatrix} \dot{y}_u \\ \dot{y}_s \end{Bmatrix}
+\begin{bmatrix} k_t + k_s & -k_s \\ -k_s & k_s \end{bmatrix}
\begin{Bmatrix} y_u \\ y_s \end{Bmatrix}
=\begin{Bmatrix} c_t\,\dot{u}_c + k_t\,u_c \\ 0 \end{Bmatrix}
}\end{equation}

where \(y_u\) and \(y_s\) are the axle and body displacements. Only the
tyre layer connects to the bridge, so the contact force is
\(F_c = k_t(y_u - u_c) + c_t(\dot{y}_u - \dot{u}_c)\).

The \textbf{half-car model with pitch} (Figure~\ref{fig-vehicle-models}
c) extends the formulation to two-axle vehicles. The body, with mass
\(m_v\) and pitch moment of inertia \(J_v\), undergoes vertical
translation \(y_v\) and rotation \(\theta_v\) about its centre of
gravity (CG). The rear and front suspensions (\(k_r\), \(c_r\) and
\(k_f\), \(c_f\)) connect the body directly to the two contact points,
without separate axle masses. Defining \(d_1\) as the distance from the
CG to the front axle and \(d_2\) as the distance from the CG to the rear
axle (Figure~\ref{fig-vehicle-models} c), and denoting the rear and
front contact-point displacements by \(u_{cr}\) and \(u_{cf}\), the
bounce equation is

\begin{equation}\phantomsection\label{eq-halfcar-bounce}{
m_v \ddot{y}_v + (c_r + c_f)\dot{y}_v + (k_r + k_f)y_v + (d_1 c_f - d_2 c_r)\dot{\theta}_v + (d_1 k_f - d_2 k_r)\theta_v = k_r u_{cr} + k_f u_{cf} + c_r \dot{u}_{cr} + c_f \dot{u}_{cf}
}\end{equation}

and the pitch equation is

\begin{equation}\phantomsection\label{eq-halfcar-pitch}{
\begin{split}
J_v \ddot{\theta}_v + (d_1 c_f - d_2 c_r)\dot{y}_v + (d_1 k_f - d_2 k_r)y_v + (d_2^2 c_r + d_1^2 c_f)\dot{\theta}_v + (d_2^2 k_r + d_1^2 k_f)\theta_v \\
= d_1 k_f u_{cf} - d_2 k_r u_{cr} + d_1 c_f \dot{u}_{cf} - d_2 c_r \dot{u}_{cr}
\end{split}
}\end{equation}

This formulation, used by Yang et al. {[}\citeproc{ref-Yang2019}{5}{]}
in their VBI element, has two dynamic DOFs and two contact points. The
rear contact force, for example, is
\(F_r = k_r(y_v - d_2\theta_v - u_{cr}) + c_r(\dot{y}_v - d_2\dot{\theta}_v - \dot{u}_{cr})\),
with an analogous expression for the front.

The \textbf{full composite half-car model}
(Figure~\ref{fig-vehicle-models} d) adds independent unsprung masses
\(m_{ur}\) and \(m_{uf}\) at each axle, each with its own tyre assembly
(\(k_{tr}\), \(c_{tr}\) and \(k_{tf}\), \(c_{tf}\)), giving four DOFs:
two axle translations, body bounce, and pitch. This model captures the
full dynamics of a two-axle truck, including the distinct frequency
ranges of tyre hop (\({\sim}10\)--15 Hz) and body bounce (\({\sim}1\)--3
Hz), and has been used by OBrien et al.
{[}\citeproc{ref-OBrien2014}{8}{]} and Cantero et al.
{[}\citeproc{ref-Cantero2024}{16}{]}.

\subsection{Coupling Methods}\label{coupling-methods}

The coupled VBI equations can be solved monolithically or through
partitioned iteration. In the monolithic approach, the vehicle DOFs are
incorporated directly into the beam element formulation; Yang's VBI
element {[}\citeproc{ref-Yang2019}{5},\citeproc{ref-Yang1997}{17}{]}
augments the standard beam element matrices with the vehicle's mass,
stiffness, and damping contributions evaluated at the current contact
position. This yields an unconditionally stable solution without
iteration but requires modification of the element library and
reassembly of the global matrices at every time step.

The partitioned approach, adopted in this work, treats the bridge and
vehicle as separate subsystems and exchanges forces and displacements
through an external interface. Green and Cebon
{[}\citeproc{ref-Green1994}{6}{]} described such a scheme for highway
bridge dynamics, and Zhu and Law {[}\citeproc{ref-Zhu2002}{7}{]}
extended it to multi-lane bridge decks. The approach is more modular
(each subsystem can use its own solver, mesh, and time step) but
requires iteration within each time step to enforce equilibrium and
compatibility at the contact points. Convergence is measured by the
relative change in the bridge displacement field between successive
iterations:

\begin{equation}\phantomsection\label{eq-convergence}{
\varepsilon^{(k)} = \frac{\sqrt{\frac{1}{N_n}\sum_{j=1}^{N_n}\left(w_j^{(k)} - w_j^{(k-1)}\right)^2}}{\max_j |w_j^{(k)}|} < \epsilon_{\text{tol}}
}\end{equation}

where \(w_j^{(k)}\) is the bridge displacement at node \(j\) in
iteration \(k\), \(N_n\) is the number of bridge nodes, and
\(\epsilon_{\text{tol}}\) is a user-specified tolerance (typically
\(10^{-4}\) to \(10^{-12}\)).

\subsection{Road Roughness}\label{road-roughness}

Road surface roughness is a primary excitation source for both the
vehicle and the bridge in VBI problems. The ISO 8608 standard
{[}\citeproc{ref-ISO8608}{18}{]} characterises road profiles through
their displacement power spectral density (PSD), which is modelled as a
function of spatial frequency:

\begin{equation}\phantomsection\label{eq-roughness-psd}{
G_d(n) = G_d(n_0) \left(\frac{n}{n_0}\right)^{-w}
}\end{equation}

where \(n\) is the spatial frequency in cycles/m, \(n_0 = 0.1\) cycles/m
is the reference frequency, \(w = 2\) is the waviness exponent, and
\(G_d(n_0)\) is the roughness coefficient that defines the road class.
The ISO 8608 standard classifies road surfaces into categories A through
E based on the value of \(G_d(n_0)\), as summarised in
Table~\ref{tbl-iso-roughness}.

\begin{longtable}[]{@{}clccc@{}}
\caption{ISO 8608 road roughness classification at reference spatial
frequency \(n_0 = 0.1\text{ cycles/m}\). Roughness coefficients
\(G_d(n_0)\) in
\(10^{-6}\text{ m}^3\).}\label{tbl-iso-roughness}\tabularnewline
\toprule\noalign{}
Class & Description & Lower limit & Geometric mean & Upper limit \\
\midrule\noalign{}
\endfirsthead
\toprule\noalign{}
Class & Description & Lower limit & Geometric mean & Upper limit \\
\midrule\noalign{}
\endhead
\bottomrule\noalign{}
\endlastfoot
A & Very good & --- & 1 & 2 \\
B & Good & 2 & 4 & 8 \\
C & Average & 8 & 16 & 32 \\
D & Poor & 32 & 64 & 128 \\
E & Very poor & 128 & 256 & 512 \\
\end{longtable}

A synthetic road roughness profile is generated as a sum of harmonics
with random phase angles:

\begin{equation}\phantomsection\label{eq-roughness-gen}{
r(x) = \sum_{i=1}^{N_h} \sqrt{2 \, G_d(n_i) \, \Delta n} \; \cos(2\pi n_i x + \phi_i)
}\end{equation}

where \(\phi_i\) are independent random phase angles uniformly
distributed over \([0, 2\pi]\), \(\Delta n\) is the spatial frequency
increment, and \(n_i\) ranges from a lower bound \(n_l\) to an upper
bound \(n_u\). In the implementation, \(n_u = 10\) cycles/m captures
wavelengths down to 0.1 m, while \(\Delta n\) is chosen such that the
period of the roughness profile (\(1/\Delta n\)) exceeds at least twice
the bridge span, preventing artificial periodicity within the analysis
window.

\subsection{Numerical Platforms for
VBI}\label{numerical-platforms-for-vbi}

VBI simulations have been implemented in a wide range of numerical
environments. Commercial finite element packages such as ABAQUS, ANSYS,
and LS-DYNA support full-vehicle and bridge models but typically require
user-written subroutines for iterative coupling, and their licensing
restricts open sharing of benchmark configurations. Research codes,
including the VBI-2D solver used to generate the NuBe dataset
{[}\citeproc{ref-CanteroVBI2D}{22}{]}, provide well-documented reference
results but are often tied to specific vehicle topologies or solver
choices. Open-source structural analysis platforms such as OpenSees
{[}\citeproc{ref-McKenna2011}{15}{]} (exposed in Python via OpenSeesPy
{[}\citeproc{ref-OpenSeesPy}{19}{]}) offer reusable building blocks for
structural dynamics but do not ship with native moving-mass analysis or
direct coupling of separate bridge and vehicle subsystems. The present
framework targets this gap: it builds on a widely used open-source
platform and adds an external Python coupling layer that works with
standard elements, so that complete VBI models and benchmark cases can
be shared, inspected, and extended without modifying the underlying
solver.

\section{Methodology}\label{sec-methodology}

\subsection{Overview}\label{overview}

\begin{figure}[htbp]

\centering{

\includegraphics[width=0.95\linewidth,height=\textheight,keepaspectratio]{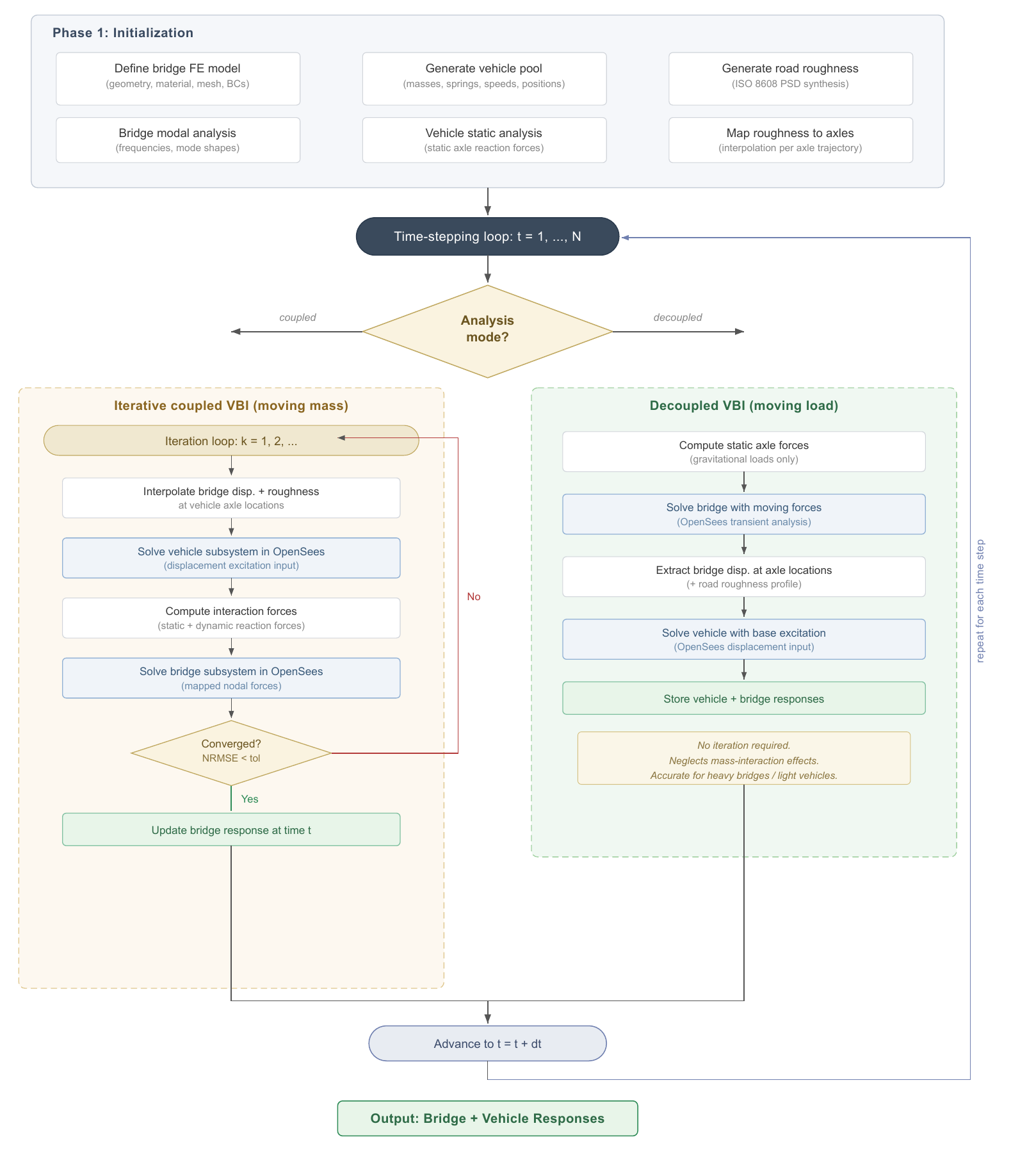}

}

\caption{\label{fig-flowchart}Framework flowchart showing the complete
analysis procedure, including initialisation, the time-stepping loop,
and the two analysis modes (coupled iterative and decoupled moving
load).}

\end{figure}

\begin{figure}[htbp]

\centering{

\includegraphics[width=0.9\linewidth,height=\textheight,keepaspectratio]{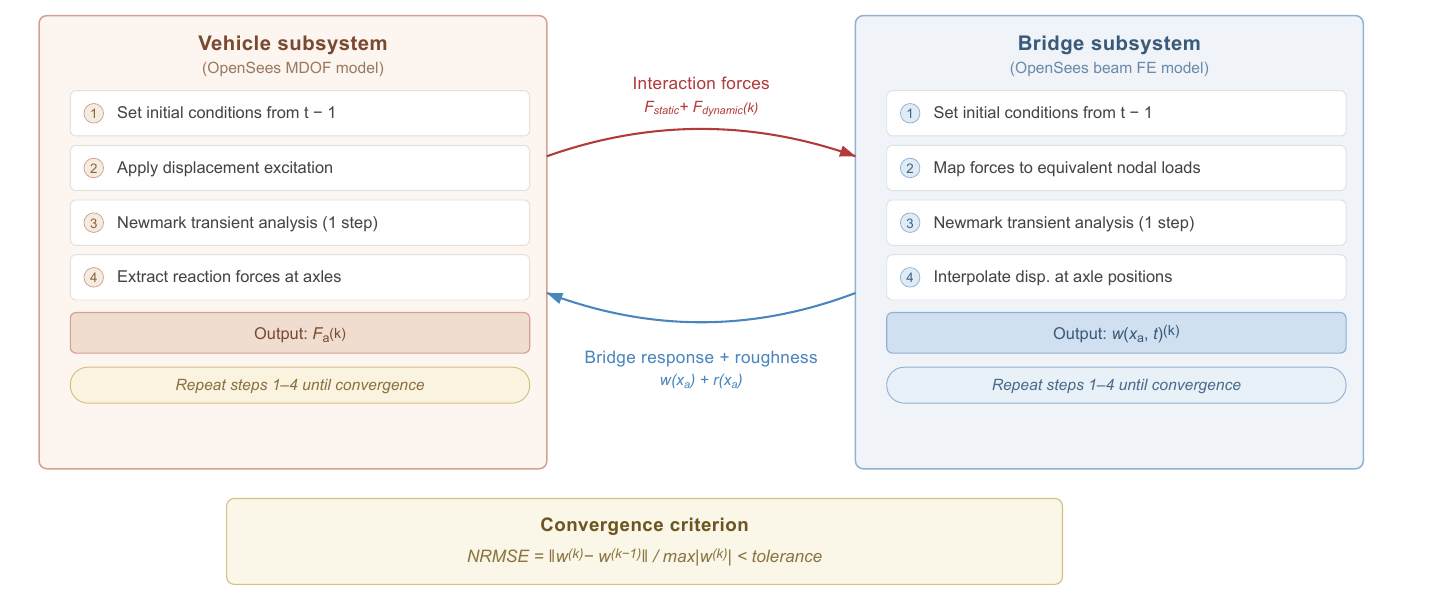}

}

\caption{\label{fig-coupling}Iterative partitioned coupling scheme at a
single time step, showing the exchange of displacement and force
quantities between the vehicle and bridge subsystems.}

\end{figure}

The proposed framework analyses the VBI problem by constructing two
independent OpenSees finite element models at each time step: one for
the bridge and one for the vehicle. The bridge model is a standard
Euler-Bernoulli beam discretised with \texttt{elasticBeamColumn}
elements, while the vehicle model uses lumped masses and
\texttt{zeroLength} spring-dashpot elements to represent the suspension
and tyre assemblies. At each time step, the two models are coupled
through the following iterative procedure: (i) the bridge deformation at
the vehicle axle locations is interpolated and combined with the road
roughness to form the vehicle input; (ii) the vehicle model is solved
for one time step under this displacement excitation; (iii) the vehicle
reaction forces are mapped back onto the bridge nodes; (iv) the bridge
model is solved for one time step under the updated nodal forces; and
(v) convergence is checked against the bridge displacement. This process
is illustrated schematically in Figure~\ref{fig-flowchart} and
Figure~\ref{fig-coupling} (see also Figure~\ref{fig-vbi-schematic} for
the physical problem setup).

\subsection{Bridge Finite Element
Model}\label{bridge-finite-element-model}

The bridge is modelled as a two-dimensional Euler-Bernoulli beam using
OpenSees \texttt{elasticBeamColumn} elements in a 2D frame formulation
with three degrees of freedom per node (two translations and one
rotation). The bridge is discretised into \(N_e\) elements of uniform
length \(\Delta x = L / N_e\), producing \(N_n = N_e + 1\) nodes. Lumped
mass is assigned to each node proportional to the tributary length:
interior nodes receive \(\bar{m} \Delta x\) and the two end nodes
receive \(\bar{m} \Delta x / 2\), where \(\bar{m}\) is the mass per unit
length. Boundary conditions are applied at the support nodes: the first
support is modelled as a pin (fixed in \(x\) and \(y\), free in
rotation), and subsequent supports are modelled as rollers (fixed in
\(y\) only). Multi-span continuous bridges are supported by specifying
the locations of intermediate supports.

Rayleigh damping is applied using the mass-proportional and
committed-stiffness-proportional coefficients computed from the target
damping ratio \(\zeta\) and the first two natural frequencies:

\begin{equation}\phantomsection\label{eq-rayleigh}{
\alpha_M = \zeta \, \frac{2 \omega_1 \omega_2}{\omega_1 + \omega_2}, \qquad \beta_K = \frac{2\zeta}{\omega_1 + \omega_2}
}\end{equation}

where \(\omega_1\) and \(\omega_2\) are the first two circular natural
frequencies obtained from an eigenvalue analysis of the bridge model.
Time integration is performed using the Newmark method
{[}\citeproc{ref-Newmark1959}{20}{]} with parameters \(\gamma = 0.5\)
and \(\beta = 0.25\) (constant average acceleration), which provides
unconditional stability {[}\citeproc{ref-Chopra2017}{21}{]}.

At each time step, the bridge model is rebuilt from scratch in OpenSees
(after a \texttt{wipe} command), the initial conditions from the
previous step are imposed using \texttt{setNodeDisp},
\texttt{setNodeVel}, and \texttt{setNodeAccel}, the nodal force time
series is applied, and a single transient analysis step is performed.
This rebuild-per-step strategy eliminates the need to maintain
persistent state within OpenSees and enables straightforward
implementation of the iterative coupling without conflicts between
vehicle and bridge model definitions.

\subsection{Vehicle Finite Element
Models}\label{vehicle-finite-element-models}

The framework implements the four vehicle model types described in
Section~\ref{sec-literature} (see Figure~\ref{fig-vehicle-models}). All
vehicle models are constructed in OpenSees using lumped masses at the
DOF nodes and \texttt{zeroLength} or \texttt{twoNodeLink} elements with
elastic materials to represent springs and dashpots. For all vehicle
models, the axle (wheel) nodes are constrained as fixed supports in the
OpenSees model. At each time step, both the displacement and velocity at
each axle are prescribed using OpenSees' \texttt{MultipleSupport}
pattern with \texttt{imposedMotion}, where the displacement time series
is the sum of the interpolated bridge deformation and the road roughness
at the axle location, and the velocity time series is the corresponding
time derivative. This combined displacement--velocity excitation ensures
that the Newmark integrator receives consistent kinematic input,
avoiding spurious accelerations that would arise from differencing
displacement alone. The reaction forces at the axle nodes, obtained via
OpenSees' \texttt{nodeReaction} command with the \texttt{-dynamic} and
\texttt{-rayleigh} flags, represent the total dynamic contact force
between the vehicle and the bridge.

For the quarter-car models (simple and composite), the OpenSees model is
constructed in one dimension (\texttt{ndm=1}), with the axle node fixed
and the body and axle masses connected by \texttt{zeroLength}
spring-dashpot elements. For the half-car models with body pitch (comp2
and comp3), a 2D frame formulation (\texttt{ndm=2}, \texttt{ndf=3}) is
required to represent the rotational DOF. The body centre of gravity is
modelled as a node carrying the body mass and the pitch moment of
inertia, connected to the suspension attachment points above each axle
through very stiff penalty elements (\texttt{twoNodeLink} with stiffness
\(\sim\!10^6 \times k_s\)). These penalty links enforce near-rigid-body
kinematics between the CG node and the suspension tops, so that body
bounce and pitch are transmitted to the axle positions through the
suspension springs without introducing additional flexible modes.
Table~\ref{tbl-vehicle-topology} in the Appendix gives the complete node
and element connectivity for each vehicle model type.

A static analysis is performed once at the beginning of the simulation
to determine the static axle forces under gravity. These forces
represent the baseline moving load that acts on the bridge even in the
absence of dynamic effects.

\subsection{Force Mapping}\label{force-mapping}

The vehicle contact forces must be mapped from the axle positions (which
generally do not coincide with bridge nodes) onto the nearest bridge
nodes. Linear shape function interpolation is used: for an axle located
at position \(x_a\) within element \([x_i, x_{i+1}]\) of length
\(\ell\), the force \(F_a\) is distributed to the two element nodes as

\begin{equation}\phantomsection\label{eq-force-mapping}{
F_i = \left(1 - \frac{x_a - x_i}{\ell}\right) F_a, \qquad F_{i+1} = \frac{x_a - x_i}{\ell} \, F_a
}\end{equation}

where \(x_i\) and \(x_{i+1}\) are the coordinates of the element end
nodes. This linear interpolation is consistent with the linear
(Hermitian) displacement interpolation within a beam element and ensures
that the total force and its first moment are preserved. When an axle is
located outside the bridge span (i.e., before entering or after exiting
the bridge), its force contribution is excluded from the bridge loading.

\subsection{Iterative Coupling
Algorithm}\label{iterative-coupling-algorithm}

The complete iterative coupling procedure at time step \(t_{n+1}\) is
formalised in Algorithm 1.

\begin{algorithm}[H]
\caption{Iterative Coupled VBI Analysis at Time Step $t_{n+1}$}
\label{alg:iterative}
\begin{algorithmic}[1]
\Require Bridge state at $t_n$: $\{\mathbf{w}_n, \dot{\mathbf{w}}_n, \ddot{\mathbf{w}}_n\}$; Vehicle state at $t_n$; Road roughness $r(x)$
\Ensure Updated bridge and vehicle states at $t_{n+1}$
\State Initialise $\mathbf{w}^{(0)}_{n+1}$ from the free-vibration bridge response (no vehicle forces)
\State Set iteration counter $k \gets 0$
\Repeat
    \State $k \gets k + 1$
    \For{each vehicle $v$}
        \For{each axle $i$ of vehicle $v$}
            \State Compute axle position: $x_{a,i} = x_{\text{rear}}(t_{n+1}) + d_i$
            \State Interpolate bridge displacement and velocity at $x_{a,i}$
            \State Compute axle displacement: $u_i = w_b(x_{a,i}) + r(x_{a,i})$
            \State Compute axle velocity: $\dot{u}_i = \dot{w}_b(x_{a,i}) + \dot{r}(x_{a,i})$
        \EndFor
        \State Solve vehicle FE model with imposed displacement $\{u_i\}$ and velocity $\{\dot{u}_i\}$
        \State Extract axle reaction forces $\{F_{a,i}\}$ via \texttt{nodeReaction(-dynamic, -rayleigh)}
    \EndFor
    \State Initialise nodal force vector: $\mathbf{F}_{\text{inter}} \gets \mathbf{0}$
    \For{each vehicle $v$, each axle $i$}
        \State $F_{\text{total},i} \gets F_{\text{static},i} + F_{a,i}^{\text{dynamic}}$
        \If{$F_{\text{total},i} > 0$} \Comment{Check for wheel uplift}
            \State Map $F_{\text{total},i}$ to bridge nodes using @eq-force-mapping
        \EndIf
    \EndFor
    \State Add background traffic force: $\mathbf{F}_n \gets \mathbf{F}_{\text{traffic}} + \mathbf{F}_{\text{inter}}$
    \State Solve bridge FE model in OpenSees with nodal forces $\mathbf{F}_n$
    \State Extract updated bridge response $\mathbf{w}^{(k)}_{n+1}$
    \State Compute convergence residual $\varepsilon^{(k)}$ using Equation~\ref{eq-convergence}
\Until{$\varepsilon^{(k)} < \epsilon_{\text{tol}}$}
\State Store converged bridge and vehicle states at $t_{n+1}$
\end{algorithmic}
\end{algorithm}

The convergence tolerance on the bridge displacement residual
$\varepsilon^{(k)}$ (Equation~\ref{eq-convergence}) is set to
\(\epsilon_{\text{tol}} = 10^{-12}\) for all validation studies in this
paper. In practice, values of \(10^{-4}\) to \(10^{-6}\) are sufficient
for engineering accuracy. The algorithm typically converges in one to
four iterations per time step for typical vehicle-to-bridge mass ratios.

\subsection{Simplified Decoupled
Approach}\label{simplified-decoupled-approach}

For situations where the vehicle mass is small relative to the bridge
mass, or when computational efficiency is paramount, the framework
provides a simplified decoupled (non-iterative) analysis mode. This
approach is summarised in Algorithm 2.

\begin{algorithm}[H]
\caption{Decoupled VBI Analysis (Moving Load + Base Excitation)}
\label{alg:decoupled}
\begin{algorithmic}[1]
\Require Bridge properties; Vehicle properties; Road roughness $r(x)$; Vehicle trajectory $x(t)$
\Ensure Bridge and vehicle response histories
\State Compute vehicle static axle forces $\{F_{\text{static},i}\}$
\For{each time step $t_n$}
    \State Compute axle positions $\{x_{a,i}(t_n)\}$
    \State Map static forces to bridge nodes using @eq-force-mapping
\EndFor
\State Solve bridge FE model for all time steps under mapped static forces
\State Extract bridge displacement history at all nodes: $\mathbf{w}(x, t)$
\For{each vehicle $v$, each axle $i$, each time step $t_n$}
    \State Interpolate bridge displacement and velocity at axle position $x_{a,i}(t_n)$
    \State $u_i(t_n) = w_b(x_{a,i}, t_n) + r(x_{a,i}(t_n))$; \quad $\dot{u}_i(t_n) = \dot{w}_b(x_{a,i}, t_n) + \dot{r}(x_{a,i}(t_n))$
\EndFor
\State Solve vehicle FE model with imposed displacement and velocity histories $\{u_i(t), \dot{u}_i(t)\}$
\end{algorithmic}
\end{algorithm}

The decoupled approach neglects the feedback of the vehicle's dynamic
forces on the bridge, treating the vehicle purely as a constant moving
load. This is equivalent to ignoring the inertial coupling between the
vehicle and bridge subsystems. The approach is exact in the limit of
zero vehicle mass (moving force problem) and becomes increasingly
approximate as the vehicle-to-bridge mass ratio increases. Its principal
advantage is computational efficiency: the bridge needs to be solved
only once (rather than iteratively at each time step), and the vehicle
can also be solved in a single pass using the precomputed bridge
response as input.

\subsection{Road Roughness Generation}\label{road-roughness-generation}

Road roughness profiles are generated according to the ISO 8608 spectral
model described in Equation~\ref{eq-roughness-psd} and
Equation~\ref{eq-roughness-gen}. The spatial frequency range is set from
\(n_l\) to \(n_u\) with increment \(\Delta n\). The generated profile is
optionally smoothed with a moving average filter to attenuate
unrealistically high-frequency components that arise from the finite
harmonic summation. The roughness profile extends over the full analysis
window, which includes approach roads before and after the bridge span,
ensuring that the vehicle suspension is excited by road roughness before
it enters the bridge.

The roughness input to each axle accounts for the spatial lag between
axles. For a vehicle with axles at local coordinates
\(\{d_0, d_1, \ldots, d_{N_a-1}\}\) relative to the rear axle and a rear
axle position history \(x_{\text{rear}}(t)\), the roughness input to
axle \(i\) at time \(t\) is \(r(x_{\text{rear}}(t) + d_i)\).

\section{Validation}\label{sec-validation}

This section validates the framework against established benchmark
configurations from the VBI literature, covering three levels of vehicle
model complexity and two independent reference sources.

\subsection{\texorpdfstring{Benchmark 1: Quarter-Car Vehicle
{[}\citeproc{ref-Yang2004book}{2}{]}}{Benchmark 1: Quarter-Car Vehicle {[}2{]}}}\label{sec-bench1}

The first benchmark reproduces the configuration from Yang et al.
{[}\citeproc{ref-Yang2004book}{2}{]}, who analysed a quarter-car
spring-mass vehicle traversing a simply supported beam. The vehicle and
bridge parameters are listed in Table~\ref{tbl-yang2004-params}.

\begin{longtable}[]{@{}lcrl@{}}
\caption{Bridge and vehicle parameters for Benchmark 1
{[}\citeproc{ref-Yang2004book}{2}{]}.}\label{tbl-yang2004-params}\tabularnewline
\toprule\noalign{}
Parameter & Symbol & Value & Unit \\
\midrule\noalign{}
\endfirsthead
\toprule\noalign{}
Parameter & Symbol & Value & Unit \\
\midrule\noalign{}
\endhead
\bottomrule\noalign{}
\endlastfoot
\textbf{Bridge} & & & \\
Span length & \(L\) & 25 & m \\
Elastic modulus & \(E\) & \(2.75 \times 10^{10}\) & Pa \\
Second moment of area & \(I\) & 0.12 & m\(^4\) \\
Cross-sectional area & \(A\) & 2.0 & m\(^2\) \\
Mass per unit length & \(\bar{m}\) & 4800 & kg/m \\
Damping ratio & \(\zeta\) & 0 & --- \\
\textbf{Vehicle} & & & \\
Sprung mass & \(m_v\) & 1200 & kg \\
Suspension stiffness & \(k_v\) & \(5.0 \times 10^5\) & N/m \\
Suspension damping & \(c_v\) & 0 & N s/m \\
Speed & \(v\) & 10 & m/s \\
\end{longtable}

\begin{figure}[htbp]

\centering{

\includegraphics[width=0.75\linewidth,height=\textheight,keepaspectratio]{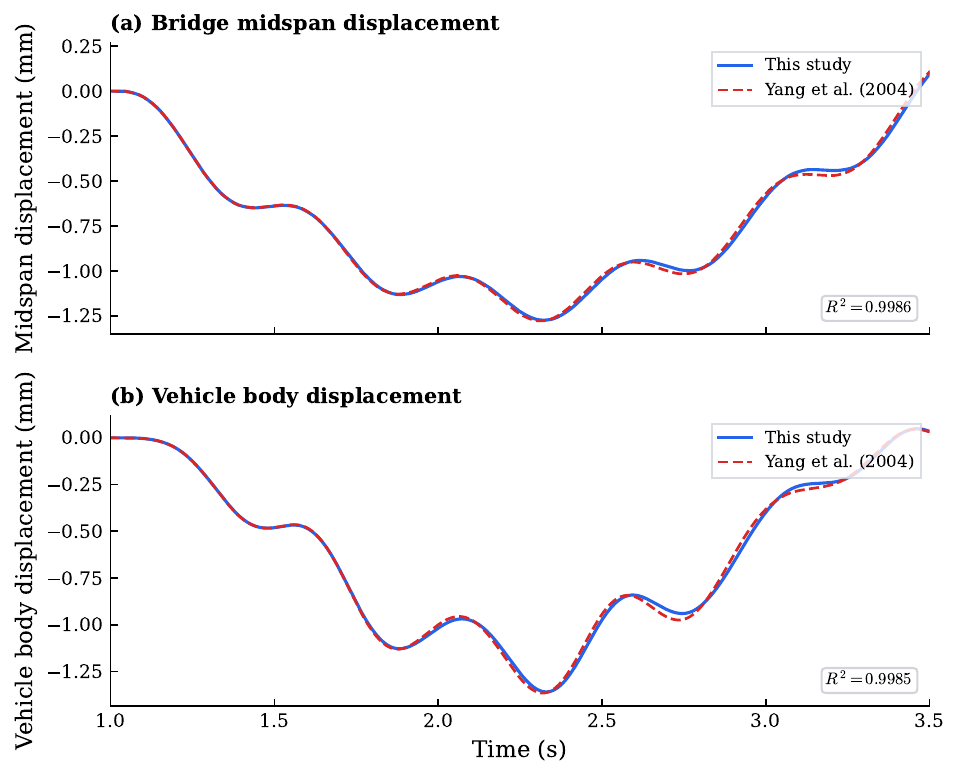}

}

\caption{\label{fig-bench1}Benchmark 1: this study (solid blue) and
Yang et al.~(2004) benchmark (dashed red) for a quarter-car vehicle
traversing a 25 m simply supported beam
{[}\citeproc{ref-Yang2004book}{2}{]}. (a) Bridge midspan displacement;
(b) vehicle body displacement.}

\end{figure}%

Figure~\ref{fig-bench1} compares this study with the Yang et al.
{[}\citeproc{ref-Yang2004book}{2}{]} quarter-car benchmark.
Agreement between the two curves is quantified using the coefficient of
determination
\begin{equation}\phantomsection\label{eq-r2}{
R^{2} = 1 - \frac{\sum_n \bigl(y_s(t_n) - y_b(t_n)\bigr)^2}{\sum_n \bigl(y_s(t_n) - \bar{y}_s\bigr)^2},
}\end{equation}
where \(y_s\) and \(y_b\) are the present-study and benchmark histories
and \(\bar{y}_s\) is the mean of \(y_s\) over the on-bridge window; $R^{2}$
approaches unity when the two signals coincide. For this benchmark,
$R^{2}$ exceeds 0.998 for both bridge midspan and vehicle body
displacement (annotated on Figure~\ref{fig-bench1}), consistent with the
visual overlap. The vehicle mass (1200 kg) is small relative to the
bridge mass (4800 kg/m \(\times\) 25 m = 120,000 kg), yielding a mass
ratio of 1\%. The bridge fundamental frequency (2.08 Hz) and the
vehicle frequency (3.25 Hz) are well separated, and the response is
dominated by the first beam mode.

\subsection{\texorpdfstring{Benchmark 2: Half-Car Vehicle with Pitch
{[}\citeproc{ref-Yang2019}{5}{]}}{Benchmark 2: Half-Car Vehicle with Pitch {[}5{]}}}\label{sec-bench2}

The second benchmark increases the vehicle model complexity by using the
half-car asymmetric vehicle with body pitch from Yang et al.
{[}\citeproc{ref-Yang2019}{5}{]}. The vehicle and bridge parameters are
listed in Table~\ref{tbl-yang2019-params}.

\begin{longtable}[]{@{}
  >{\raggedright\arraybackslash}p{(\linewidth - 6\tabcolsep) * \real{0.3889}}
  >{\centering\arraybackslash}p{(\linewidth - 6\tabcolsep) * \real{0.1528}}
  >{\raggedleft\arraybackslash}p{(\linewidth - 6\tabcolsep) * \real{0.3194}}
  >{\raggedright\arraybackslash}p{(\linewidth - 6\tabcolsep) * \real{0.1389}}@{}}
\caption{Bridge and vehicle parameters for Benchmark 2
{[}\citeproc{ref-Yang2019}{5}{]}.}\label{tbl-yang2019-params}\tabularnewline
\toprule\noalign{}
\begin{minipage}[b]{\linewidth}\raggedright
Parameter
\end{minipage} & \begin{minipage}[b]{\linewidth}\centering
Symbol
\end{minipage} & \begin{minipage}[b]{\linewidth}\raggedleft
Value
\end{minipage} & \begin{minipage}[b]{\linewidth}\raggedright
Unit
\end{minipage} \\
\midrule\noalign{}
\endfirsthead
\toprule\noalign{}
\begin{minipage}[b]{\linewidth}\raggedright
Parameter
\end{minipage} & \begin{minipage}[b]{\linewidth}\centering
Symbol
\end{minipage} & \begin{minipage}[b]{\linewidth}\raggedleft
Value
\end{minipage} & \begin{minipage}[b]{\linewidth}\raggedright
Unit
\end{minipage} \\
\midrule\noalign{}
\endhead
\bottomrule\noalign{}
\endlastfoot
\textbf{Bridge} & & & \\
Span length & \(L\) & 30 & m \\
Flexural rigidity & \(EI\) & \(1.56 \times 10^{10}\) & N m\(^2\) \\
Mass per unit length & \(\bar{m}\) & 4400 & kg/m \\
Damping ratio & \(\zeta\) & 0 & --- \\
\textbf{Vehicle} & & & \\
Body mass & \(m_v\) & 2500 & kg \\
Pitch moment of inertia & \(J_v\) & 2300 & kg m\(^2\) \\
Front suspension stiffness & \(k_f\) & \(2.3 \times 10^5\) & N/m \\
Rear suspension stiffness & \(k_r\) & \(1.8 \times 10^5\) & N/m \\
Front suspension damping & \(c_f\) & 0 & N s/m \\
Rear suspension damping & \(c_r\) & 0 & N s/m \\
Axle spacing & \(d\) & 3 & m \\
CG to rear axle & \(d_1\) & 1.7 & m \\
Speed & \(v\) & 10 & m/s \\
\end{longtable}

\begin{figure}[htbp]

\centering{

\includegraphics[width=0.75\linewidth,height=\textheight,keepaspectratio]{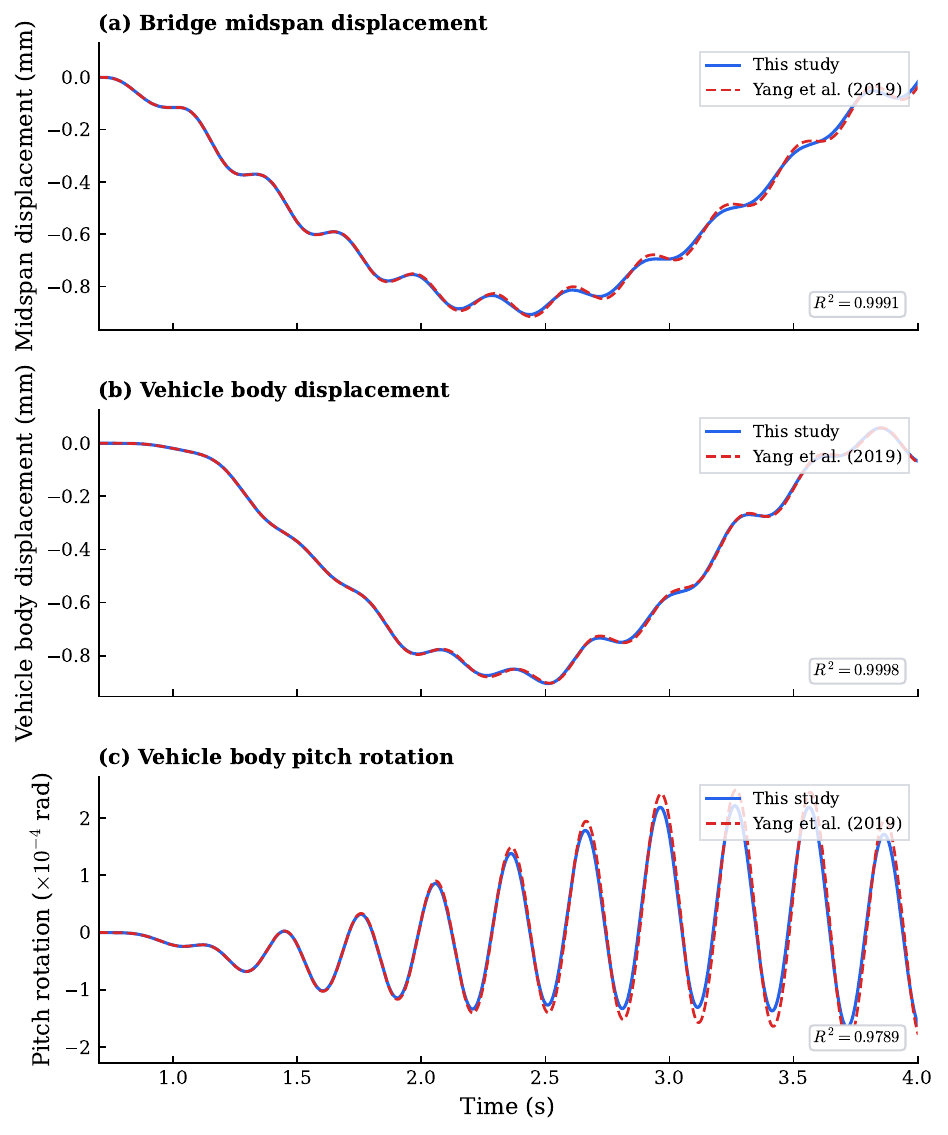}

}

\caption{\label{fig-bench2}Benchmark 2: this study (solid blue) and
Yang et al.~(2019) benchmark (dashed red) for a half-car vehicle with
pitch traversing a 30 m simply supported beam
{[}\citeproc{ref-Yang2019}{5}{]}. (a) Bridge midspan displacement; (b)
vehicle body displacement; (c) vehicle body pitch rotation.}

\end{figure}%

Figure~\ref{fig-bench2} compares this study with the Yang et al.
{[}\citeproc{ref-Yang2019}{5}{]} half-car benchmark with body pitch. The
vehicle mass (2500 kg) relative to the bridge mass (4400 kg/m
\(\times\) 30 m = 132,000 kg) yields a mass ratio of approximately
1.9\%. The maximum midspan displacement of approximately 0.85 mm occurs
when the vehicle is near the centre of the span, and the response
exhibits a multi-modal character due to the asymmetric half-car loading.
The vehicle body displacement and pitch rotation closely follow the Yang
et al.~(2019) benchmark, including near pitch-response peaks. The pitch
rotation grows in amplitude as the vehicle traverses the span,
reflecting the increasing excitation of the pitching mode by the offset
centre of gravity.

\subsection{\texorpdfstring{Benchmark 3: NuBe Vehicle Models
{[}\citeproc{ref-Cantero2024}{16}{]}}{Benchmark 3: NuBe Vehicle Models {[}16{]}}}\label{sec-bench3}

The third benchmark uses vehicle and bridge configurations from the NuBe
numerical benchmark dataset {[}\citeproc{ref-Cantero2024}{16}{]}, which
provides a standardised set of VBI simulation results generated using
the independent VBI-2D MATLAB solver
{[}\citeproc{ref-CanteroVBI2D}{22}{]}. We configure our framework with
the mean vehicle properties from the NuBe V1 (quarter-car) and V2
(half-car with pitch) models and run them on the NuBe B27 bridge (27 m
span). The bridge and vehicle parameters are listed in
Table~\ref{tbl-cantero-params}.

\begin{longtable}[]{@{}
  >{\raggedright\arraybackslash}p{(\linewidth - 6\tabcolsep) * \real{0.3896}}
  >{\centering\arraybackslash}p{(\linewidth - 6\tabcolsep) * \real{0.1558}}
  >{\raggedleft\arraybackslash}p{(\linewidth - 6\tabcolsep) * \real{0.3247}}
  >{\raggedright\arraybackslash}p{(\linewidth - 6\tabcolsep) * \real{0.1299}}@{}}
\caption{Bridge and vehicle parameters for Benchmark 3, using mean
values from the NuBe dataset
{[}\citeproc{ref-Cantero2024}{16}{]}.}\label{tbl-cantero-params}\tabularnewline
\toprule\noalign{}
\begin{minipage}[b]{\linewidth}\raggedright
Parameter
\end{minipage} & \begin{minipage}[b]{\linewidth}\centering
Symbol
\end{minipage} & \begin{minipage}[b]{\linewidth}\raggedleft
Value
\end{minipage} & \begin{minipage}[b]{\linewidth}\raggedright
Unit
\end{minipage} \\
\midrule\noalign{}
\endfirsthead
\toprule\noalign{}
\begin{minipage}[b]{\linewidth}\raggedright
Parameter
\end{minipage} & \begin{minipage}[b]{\linewidth}\centering
Symbol
\end{minipage} & \begin{minipage}[b]{\linewidth}\raggedleft
Value
\end{minipage} & \begin{minipage}[b]{\linewidth}\raggedright
Unit
\end{minipage} \\
\midrule\noalign{}
\endhead
\bottomrule\noalign{}
\endlastfoot
\textbf{Bridge (B27)} & & & \\
Span length & \(L\) & 27 & m \\
Young's modulus & \(E\) & \(3.5 \times 10^{10}\) & Pa \\
Second moment of area & \(I\) & 1.7055 & m\(^4\) \\
Mass per unit length & \(\bar{m}\) & 19,372 & kg/m \\
Damping ratio & \(\zeta\) & 0 & --- \\
First natural frequency & \(f_1\) & 3.7824 & Hz \\
\textbf{Vehicle V1 (quarter-car)} & & & \\
Body mass & \(m_B\) & 8,000 & kg \\
Axle mass & \(m_G\) & 1,100 & kg \\
Suspension stiffness & \(k_S\) & \(2.0 \times 10^{6}\) & N/m \\
Suspension damping & \(c_S\) & \(4.0 \times 10^{4}\) & N s/m \\
Tyre stiffness & \(k_T\) & \(3.5 \times 10^{6}\) & N/m \\
Speed & \(v\) & 25 & m/s \\
\textbf{Vehicle V2 (half-car)} & & & \\
Body mass & \(m_{B1}\) & 10,500 & kg \\
Axle mass (each) & \(m_G\) & 900 & kg \\
Body moment of inertia & \(I_{B1}\) & 50,000 & kg m\(^2\) \\
Suspension stiffness (each) & \(k_S\) & \(6.0 \times 10^{6}\) & N/m \\
Suspension damping (each) & \(c_S\) & \(1.0 \times 10^{4}\) & N s/m \\
Tyre stiffness (each) & \(k_T\) & \(1.75 \times 10^{6}\) & N/m \\
Axle spacing & \(d\) & 5.0 & m \\
Speed & \(v\) & 25 & m/s \\
\end{longtable}

\begin{figure}[htbp]

\centering{

\includegraphics[width=0.75\linewidth,height=\textheight,keepaspectratio]{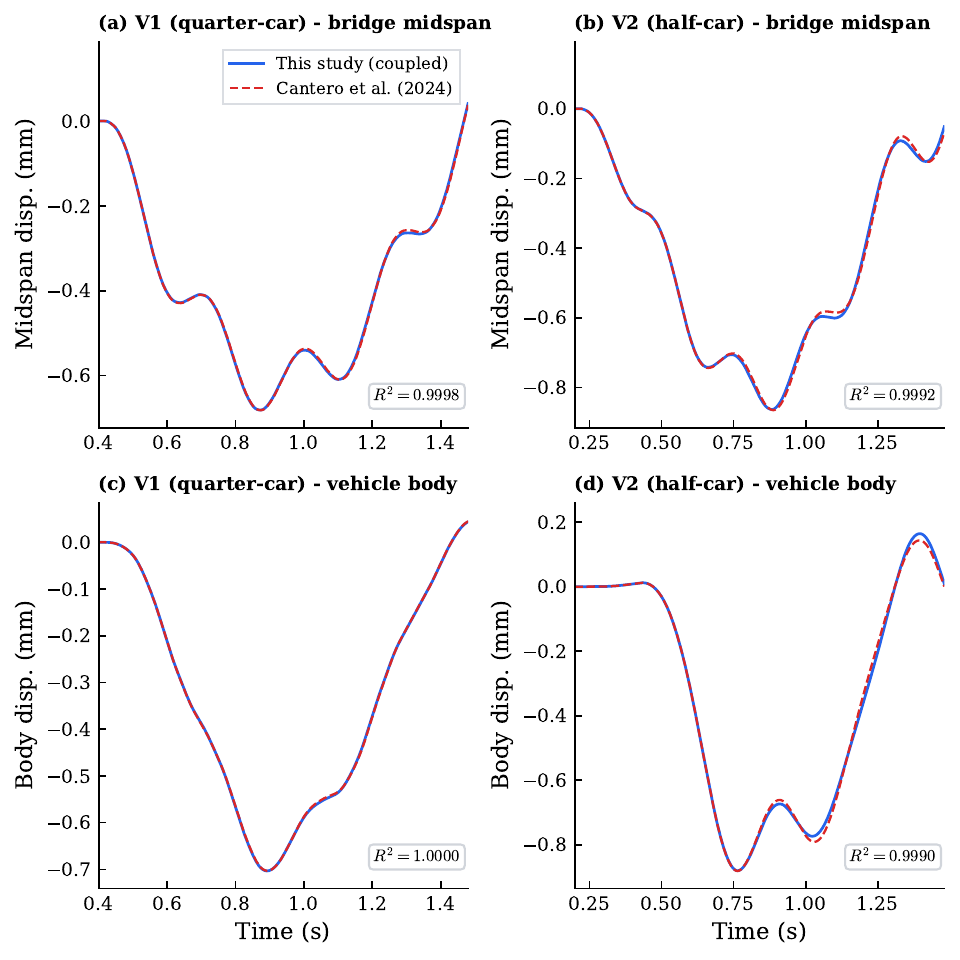}

}

\caption{\label{fig-bench3}Benchmark 3: bridge midspan displacement and
vehicle body response for the NuBe V1 (quarter-car) and V2 (half-car)
models on the B27 bridge (27 m, undamped, smooth road). Solid lines:
coupled iterative analysis; dashed lines: decoupled moving load
approach.}

\end{figure}%

The NuBe benchmark exercises two vehicle model types on the same bridge,
testing both the single-axle composite (quarter-car) and the full
two-axle composite (half-car with pitch and independent axle masses).
The V1 quarter-car model (total mass 9,100 kg, mass ratio
\(\mu \approx 1.7\%\)) produces a smooth midspan deflection dominated by
the first beam mode. The V2 half-car model (total mass 12,300 kg,
\(\mu \approx 2.4\%\)) introduces a richer response due to the pitching
motion and the separate excitation of two contact points. For both
vehicles, the coupled and decoupled analyses agree closely, consistent
with the low mass ratios. The ability to reproduce the NuBe
configurations using standard vehicle parameters from an independent
benchmark dataset {[}\citeproc{ref-Cantero2024}{16}{]} confirms the
interoperability of the proposed framework with established VBI
simulation tools.

\subsection{Benchmark 4: Convergence and Iteration
Behaviour}\label{sec-bench4}

The convergence behaviour of the iterative algorithm is examined for the
Yang et al.~(2019) half-car configuration
{[}\citeproc{ref-Yang2019}{5}{]} from Section~\ref{sec-bench2}.

\begin{figure}[htbp]

\centering{

\includegraphics[width=0.75\linewidth,height=\textheight,keepaspectratio]{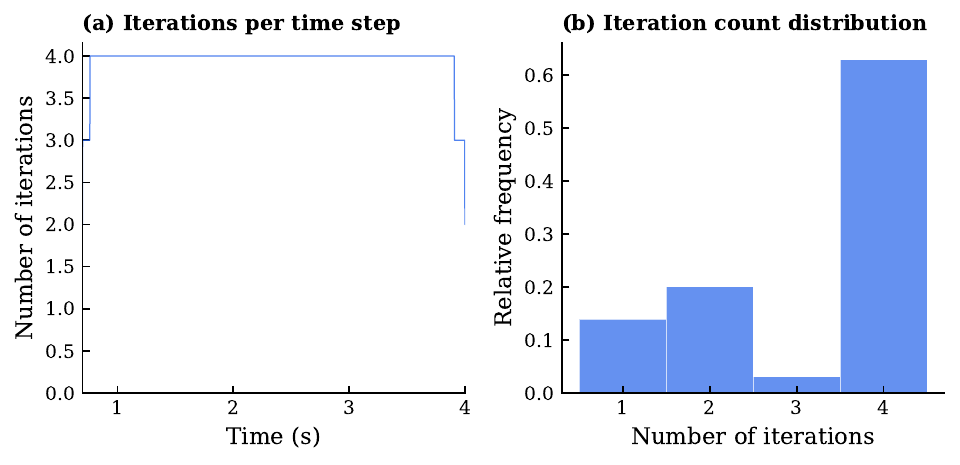}

}

\caption{\label{fig-convergence}Convergence characteristics of the
iterative coupling algorithm: (a) number of iterations per time step for
the Yang et al.~(2019) configuration; (b) iteration count distribution
showing convergence in 1--4 iterations per time step.}

\end{figure}%

The iteration count ranges from 1 to 4 across the time steps, with the
majority of steps (approximately 60\%) requiring 4 iterations and most
of the remainder converging in 1--2 iterations at the tight tolerance of
\(10^{-12}\). The iteration count is highest during the on-bridge
portion and drops as the vehicle exits. This rapid convergence confirms
the effectiveness of the iterative partitioned scheme for typical VBI
problems.

\section{Parametric Study: Coupled vs.~Decoupled
Analysis}\label{sec-parametric}

This section quantifies the conditions under which the computationally
efficient decoupled approach provides acceptable accuracy compared to
the fully coupled iterative analysis. The key parameter governing the
accuracy of the decoupled approach is the \textbf{vehicle-to-bridge mass
ratio} \(\mu = m_v / (\bar{m} L)\), where \(m_v\) is the total vehicle
mass and \(\bar{m} L\) is the total bridge mass. When \(\mu \ll 1\), the
vehicle's inertial effect on the bridge is negligible and the decoupled
approach is accurate; as \(\mu\) increases, the interaction becomes
significant and the coupled approach is necessary. The study considers
four simply supported bridge spans with realistic cross-section
properties, two quarter-car vehicle models spanning more than an order
of magnitude in mass, six vehicle speeds, three ISO 8608 roughness
classes, and varying levels of background traffic. The bridge and
vehicle configurations follow those of Eshkevari et al.
{[}\citeproc{ref-Eshkevari2020}{12}{]}.

Table~\ref{tbl-param-bridges} lists the bridge configurations used in
this section. The second moment of area \(I\) is computed from the
reported box cross-section dimensions, and the mass per unit length
\(\bar{m}\) is calibrated to match the fundamental frequency of each
span using the Euler-Bernoulli simply supported beam formula
\(f_1 = (\pi / 2L^2)\sqrt{EI/\bar{m}}\) with \(E = 27.5\) GPa. The
resulting \(\bar{m}\) values represent the equivalent distributed mass
of the full bridge deck, including slab, parapets, and wearing surface.
Two quarter-car vehicle models are used: the commercial vehicle (total
mass 535.9 kg) and the heavy truck (total mass 18,000 kg) from Eshkevari
et al. {[}\citeproc{ref-Eshkevari2020}{12}{]}. All analyses use ISO 8608
Class C road roughness, 0.3\% Rayleigh damping, and a constant vehicle
speed of 10 m/s. The spatial frequency increment for roughness
generation is set to \(\Delta n = \min(0.01, 1/(2L))\) cycles/m to
ensure that the fundamental period of the synthesised profile exceeds
twice the bridge length, avoiding artificial periodicity for longer
spans.

\begin{longtable}[]{@{}
  >{\centering\arraybackslash}p{(\linewidth - 18\tabcolsep) * \real{0.1067}}
  >{\centering\arraybackslash}p{(\linewidth - 18\tabcolsep) * \real{0.0933}}
  >{\centering\arraybackslash}p{(\linewidth - 18\tabcolsep) * \real{0.0933}}
  >{\centering\arraybackslash}p{(\linewidth - 18\tabcolsep) * \real{0.0933}}
  >{\centering\arraybackslash}p{(\linewidth - 18\tabcolsep) * \real{0.0933}}
  >{\centering\arraybackslash}p{(\linewidth - 18\tabcolsep) * \real{0.1067}}
  >{\centering\arraybackslash}p{(\linewidth - 18\tabcolsep) * \real{0.1067}}
  >{\centering\arraybackslash}p{(\linewidth - 18\tabcolsep) * \real{0.0933}}
  >{\centering\arraybackslash}p{(\linewidth - 18\tabcolsep) * \real{0.1067}}
  >{\centering\arraybackslash}p{(\linewidth - 18\tabcolsep) * \real{0.1067}}@{}}

\caption{\label{tbl-param-bridges}Bridge configurations for the
parametric study, adapted from Eshkevari et al.~The second moment of
area \(I\) is computed from the box cross-section dimensions; the mass
per unit length \(\bar{m}\) is calibrated to match the target
fundamental frequency.}

\tabularnewline

\toprule\noalign{}
\begin{minipage}[b]{\linewidth}\centering
Span (m)
\end{minipage} & \begin{minipage}[b]{\linewidth}\centering
Depth (m)
\end{minipage} & \begin{minipage}[b]{\linewidth}\centering
Width (m)
\end{minipage} & \begin{minipage}[b]{\linewidth}\centering
\(t_f\) (m)
\end{minipage} & \begin{minipage}[b]{\linewidth}\centering
\(t_w\) (m)
\end{minipage} & \begin{minipage}[b]{\linewidth}\centering
\(I\) (m\(^4\))
\end{minipage} & \begin{minipage}[b]{\linewidth}\centering
\(\bar{m}\) (kg/m)
\end{minipage} & \begin{minipage}[b]{\linewidth}\centering
\(f_1\) (Hz)
\end{minipage} & \begin{minipage}[b]{\linewidth}\centering
\(\mu_{\text{comm}}\)
\end{minipage} & \begin{minipage}[b]{\linewidth}\centering
\(\mu_{\text{heavy}}\)
\end{minipage} \\
\midrule\noalign{}
\endhead
\bottomrule\noalign{}
\endlastfoot
15 & 0.60 & 0.30 & 0.04 & 0.02 & 0.0024 & 49 & 8.03 & 0.7303 & 24.530 \\
30 & 1.10 & 0.50 & 0.05 & 0.03 & 0.0188 & 119 & 3.63 & 0.1495 & 5.022 \\
50 & 1.60 & 1.30 & 0.10 & 0.05 & 0.1693 & 437 & 2.05 & 0.0245 & 0.823 \\
100 & 2.40 & 2.00 & 0.15 & 0.10 & 0.9148 & 1104 & 0.75 & 0.0049 &
0.163 \\

\end{longtable}

\subsection{Effect of Bridge Span Length}\label{sec-span-study}

The first investigation varies the bridge span from 15 m to 100 m while
keeping the vehicle, roughness, and speed parameters fixed. Both vehicle
types are tested on each bridge to simultaneously capture the effects of
span length and vehicle mass.

\begin{figure}[htbp]

\centering{

\includegraphics[width=0.75\linewidth,height=\textheight,keepaspectratio]{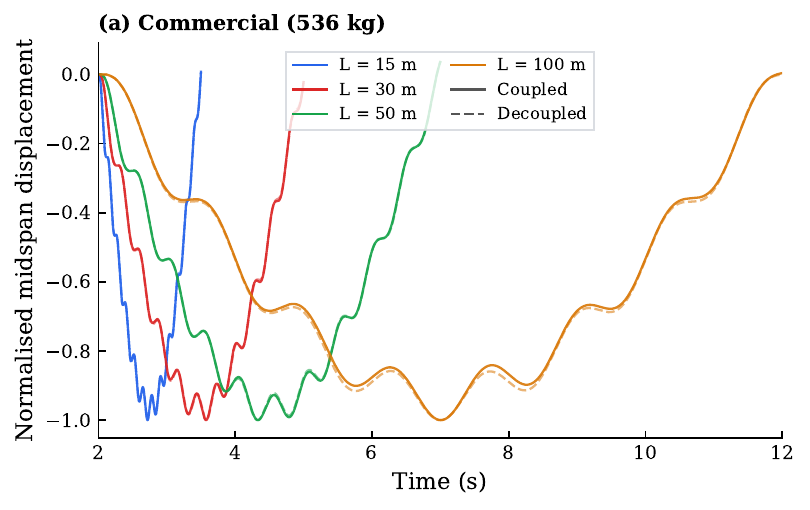}

}

\caption{\label{fig-span-a}Normalised bridge midspan displacement for
the commercial vehicle across four spans (15 m, 30 m, 50 m, 100 m).
Solid lines: coupled iterative analysis. Dashed lines: decoupled
(moving load) analysis.}

\end{figure}%

\begin{figure}[htbp]

\centering{

\includegraphics[width=0.75\linewidth,height=\textheight,keepaspectratio]{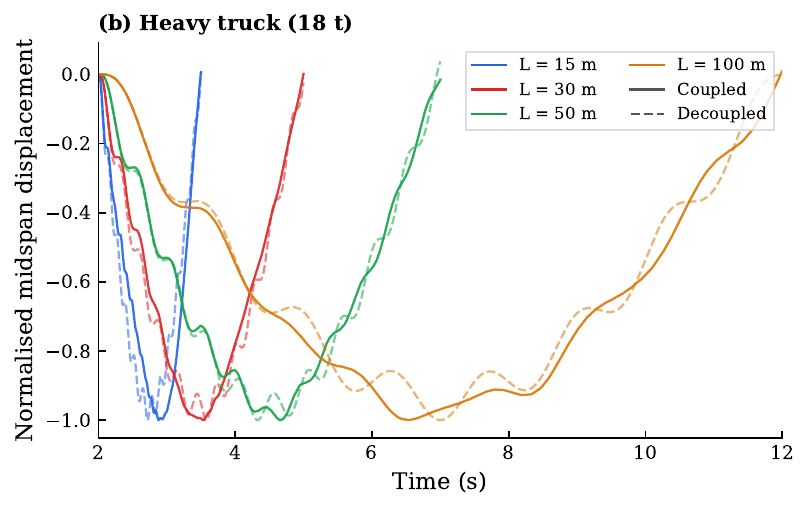}

}

\caption{\label{fig-span-b}Normalised bridge midspan displacement for
the heavy truck across four spans. Solid lines: coupled iterative
analysis. Dashed lines: decoupled (moving load) analysis.}

\end{figure}%

\begin{figure}[htbp]

\centering{

\includegraphics[width=0.75\linewidth,height=\textheight,keepaspectratio]{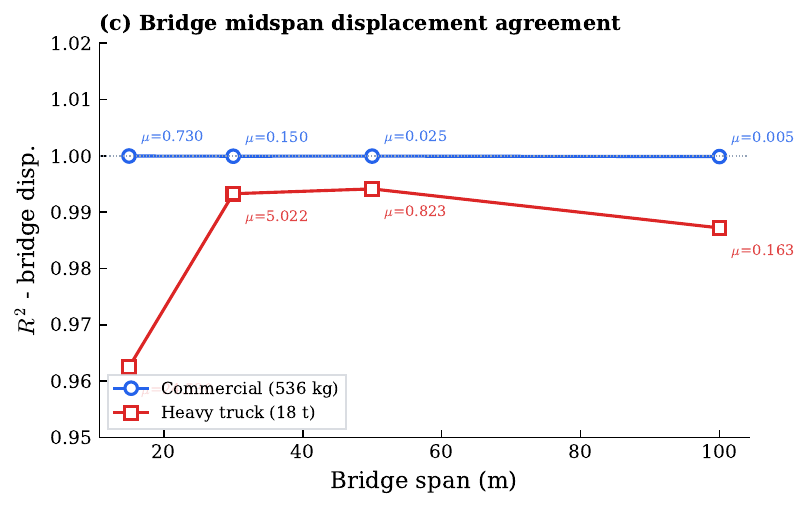}

}

\caption{\label{fig-span-c}Coefficient of determination $R^{2}$ for
bridge midspan displacement between coupled and decoupled analyses as a
function of span length, for the commercial vehicle (blue circles) and
the heavy truck (red squares). Mass ratios \(\mu\) are annotated next to
each data point. Values close to 1 indicate close agreement.}

\end{figure}%

\begin{figure}[htbp]

\centering{

\includegraphics[width=0.75\linewidth,height=\textheight,keepaspectratio]{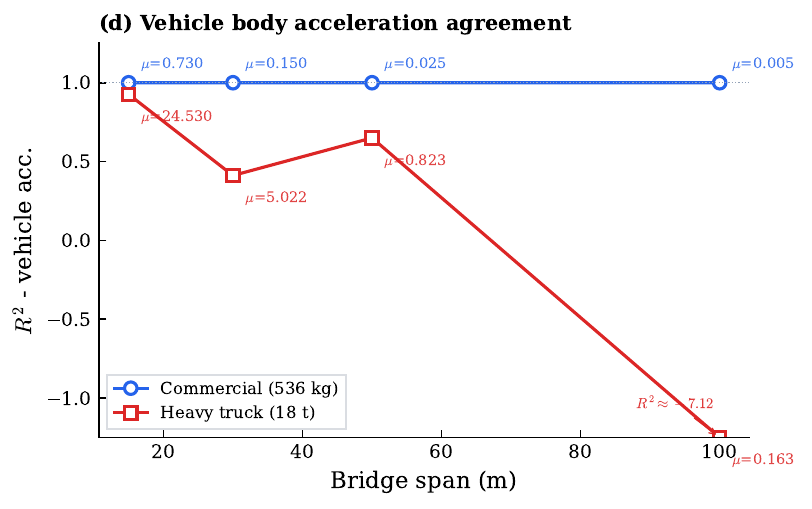}

}

\caption{\label{fig-span-d}Coefficient of determination $R^{2}$ for
vehicle body acceleration between coupled and decoupled analyses as a
function of span length, for the commercial vehicle and the heavy
truck.}

\end{figure}%

The results reveal a clear separation between the two vehicle classes.
For the commercial vehicle (\(m_v \approx 536\) kg), the coefficient of
determination $R^{2}$ between coupled and decoupled bridge displacement
remains above 0.999 across all four spans, confirming that a vehicle of
this mass has negligible inertial effect on any of the bridges tested
(Figure~\ref{fig-span-c}). For the heavy truck (\(m_v \approx 18{,}000\)
kg), the agreement deteriorates markedly: at the 15 m span
(\(\mu \approx 24.5\), where the truck outweighs the bridge), $R^{2}$
drops well below unity, and even at the 100 m span
(\(\mu \approx 0.16\)) it remains noticeably below the commercial-vehicle
value. The vehicle body acceleration $R^{2}$ (Figure~\ref{fig-span-d})
follows the same qualitative pattern with an even larger gap between the
two vehicles, reflecting the amplification of interaction effects
through the suspension dynamics.

A noteworthy feature of Figure~\ref{fig-span-d} is that the vehicle
body acceleration $R^{2}$ drops sharply and becomes strongly negative
for the heavy truck at the longest span (\(L = 100\) m,
\(\mu \approx 0.163\), $R^{2} \approx -7$), while the commercial
vehicle at a similarly small mass ratio (\(\mu \approx 0.005\) at
\(L = 100\) m) remains at $R^{2} > 0.9999$. The two data points sit at
the same end of the span axis but the two vehicle classes behave
differently. This is a well-known property of the coefficient of
determination when applied to low-variance signals, not a sign that the
decoupled approximation has suddenly collapsed. Recall that $R^{2}$
normalises the coupled-vs-decoupled residual by the variance of the
coupled signal about its mean, i.e.\ by
\(\sum_n (y_c - \bar{y}_c)^2\). At long spans the bridge fundamental
frequency is very low (0.75 Hz for \(L = 100\) m in the Eshkevari set)
and the bridge-induced vehicle body acceleration is small, so the
mean-centred variance \(\sum_n (y_c - \bar{y}_c)^2\) becomes small.
For the heavy truck this small denominator is combined with a residual
coupling effect that is still non-negligible because the truck carries
enough inertia to feed energy back into the bridge; the numerator
\(\sum_n (y_c - y_d)^2\) therefore dominates the denominator and
pushes $R^{2}$ strongly negative, indicating only that the decoupled
prediction is farther from the coupled signal than a constant equal to
the coupled mean. It does \emph{not} imply a large absolute error: the
underlying RMS difference is still a small fraction of the bridge's
static moving-load response. For the commercial vehicle, the residual
coupling effect is negligibly small because the vehicle mass is minute
compared with the bridge mass, so the numerator collapses faster than
the denominator and $R^{2}$ remains essentially unity. The same
degenerate behaviour does not appear in the bridge midspan displacement
(Figure~\ref{fig-span-c}), because bridge displacement has a dominant
quasi-static component and hence a much larger
\(\sum_n (y_c - \bar{y}_c)^2\), so the denominator never collapses.

These results demonstrate that the mass ratio \(\mu\) is the primary
predictor of decoupled accuracy: for the lightweight commercial vehicle
(\(\mu < 1\)), the decoupled approach is adequate regardless of span,
while for heavy vehicles (\(\mu > 0.1\)), coupled analysis is necessary
across all spans tested.

\subsection{Effect of Vehicle Speed}\label{sec-speed-study}

This investigation examines the effect of vehicle speed on both bridge
and vehicle responses using the 30 m bridge and the commercial
quarter-car vehicle under Class A roughness. The vehicle speed is varied
from 5 to 30 m/s (18 to 108 km/h), covering the typical range of highway
traffic.

\begin{figure}[htbp]

\centering{

\includegraphics[width=0.75\linewidth,height=\textheight,keepaspectratio]{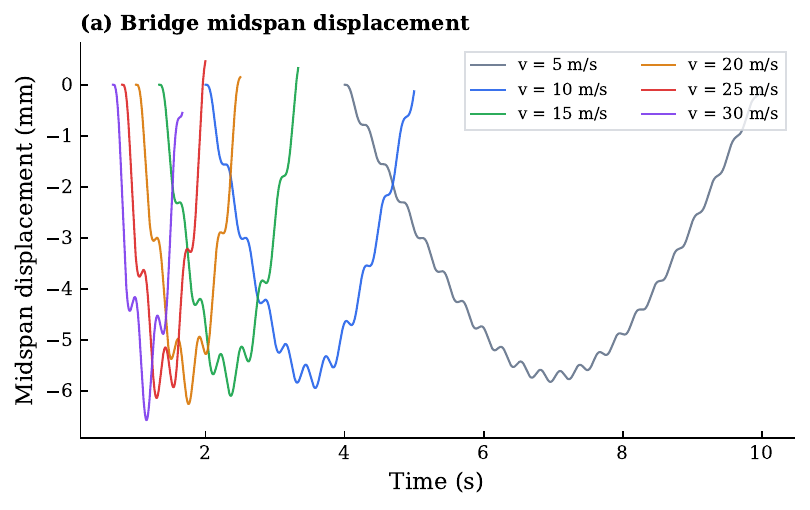}

}

\caption{\label{fig-speed-a}Bridge midspan displacement for the 30 m
bridge under the commercial quarter-car vehicle at six vehicle speeds
(5, 10, 15, 20, 25, 30 m/s), Class A roughness.}

\end{figure}%

\begin{figure}[htbp]

\centering{

\includegraphics[width=0.75\linewidth,height=\textheight,keepaspectratio]{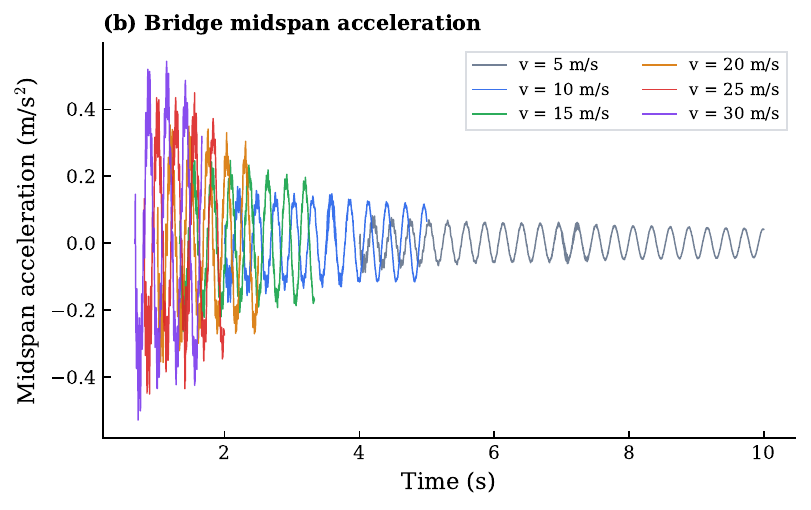}

}

\caption{\label{fig-speed-b}Bridge midspan acceleration for the same
six vehicle speeds.}

\end{figure}%

\begin{figure}[htbp]

\centering{

\includegraphics[width=0.75\linewidth,height=\textheight,keepaspectratio]{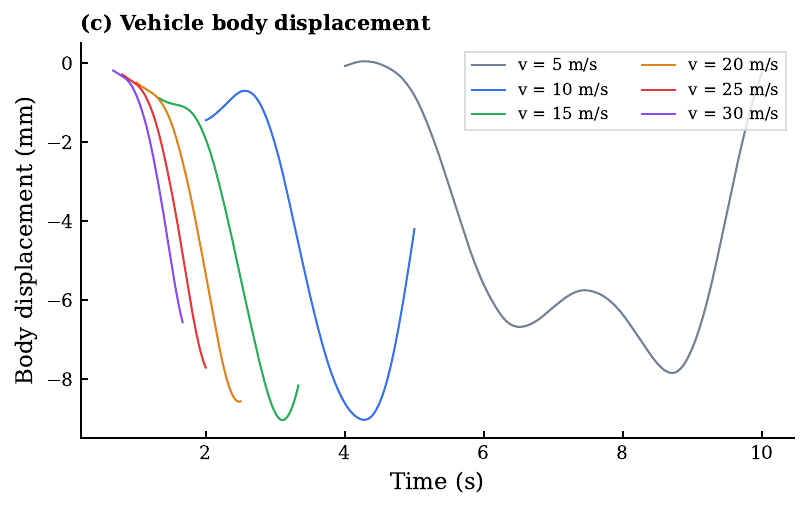}

}

\caption{\label{fig-speed-c}Vehicle body vertical displacement for the
same six vehicle speeds.}

\end{figure}%

\begin{figure}[htbp]

\centering{

\includegraphics[width=0.75\linewidth,height=\textheight,keepaspectratio]{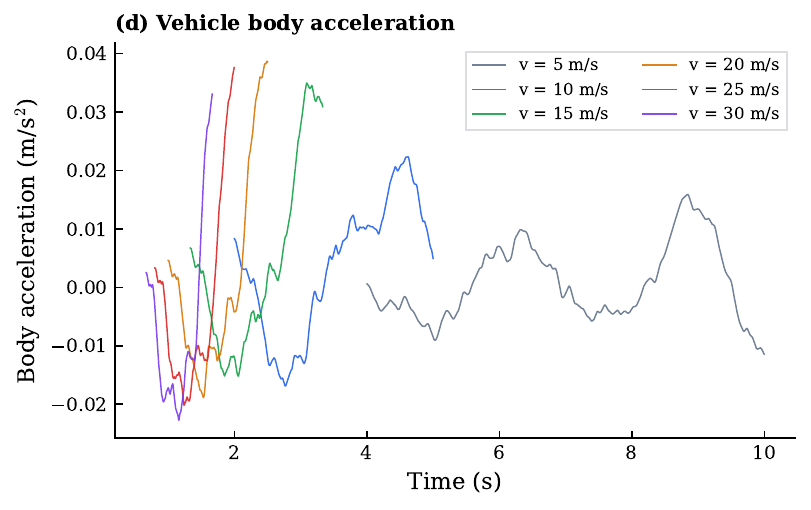}

}

\caption{\label{fig-speed-d}Vehicle body vertical acceleration for the
same six vehicle speeds.}

\end{figure}%

The vehicle speed governs the loading frequency seen by the bridge:
faster vehicles excite higher-frequency content and spend less time on
the span. At lower speeds (5--10 m/s), the bridge midspan displacement
approaches the quasi-static deflection, with the response dominated by
the first beam mode (Figure~\ref{fig-speed-a}). As speed increases,
dynamic amplification becomes apparent: the peak midspan displacement
and the free-vibration oscillations after the vehicle exits grow with
speed. The bridge midspan acceleration (Figure~\ref{fig-speed-b})
increases markedly with speed, as the moving load excites the bridge at
higher frequencies. The vehicle body displacement
(Figure~\ref{fig-speed-c}) shows a similar trend, with higher speeds
producing larger oscillations due to the more rapid traversal of the
bridge deflection profile. The vehicle body acceleration
(Figure~\ref{fig-speed-d}) also increases with speed, reflecting the
combined effect of dynamic bridge response and the higher-frequency
roughness excitation at elevated velocities.

\subsection{Road Roughness Effects}\label{sec-roughness-study}

\begin{figure}[htbp]

\centering{

\includegraphics[width=0.75\linewidth,height=\textheight,keepaspectratio]{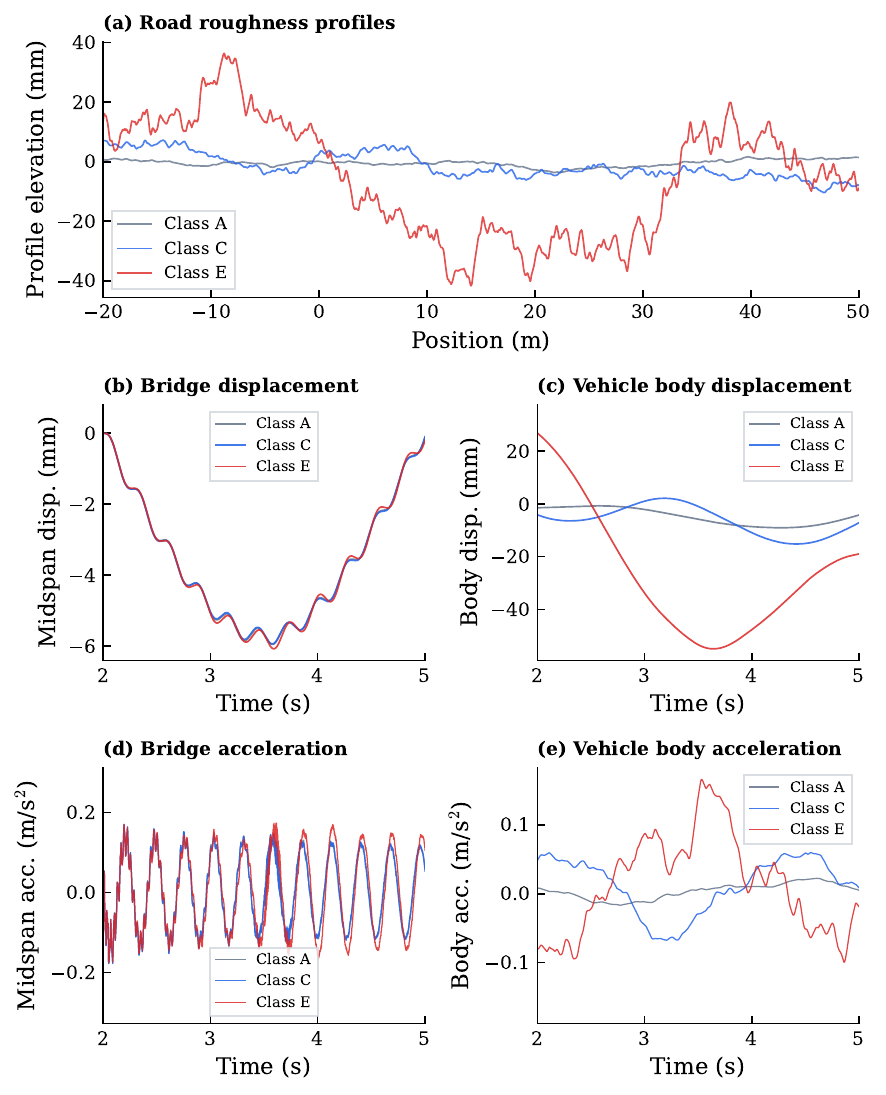}

}

\caption{\label{fig-roughness-study}Effect of road roughness on bridge
and vehicle response (30 m bridge, commercial quarter-car vehicle). (a)
ISO 8608 roughness profiles for Classes A, C, and E. (b) Bridge midspan
displacement. (c) Vehicle body vertical displacement. (d) Bridge midspan
acceleration. (e) Vehicle body vertical acceleration.}

\end{figure}%

Road roughness significantly affects the dynamic component of both
bridge and vehicle responses. Under Class A roughness, bridge
displacement and vehicle body displacement remain close to the
quasi-static deflection under the vehicle weight
(Figure~\ref{fig-roughness-study} b, c). Classes C and E introduce
progressively larger dynamic oscillations, particularly in the vehicle
body, which is directly excited by the road profile through the
suspension: the vehicle body acceleration amplitude under Class E
roughness exceeds that of Class A by over an order of magnitude
(Figure~\ref{fig-roughness-study} e). The bridge midspan acceleration
(Figure~\ref{fig-roughness-study} d) shows a comparatively smaller
sensitivity to roughness class for this lightweight vehicle, as the
bridge dynamics are dominated by its own modal response rather than the
roughness-induced contact forces. For heavier vehicles, road roughness
would play a larger role in the bridge acceleration, confirming that
surface condition is an important excitation source for indirect bridge
monitoring applications
{[}\citeproc{ref-Yang2004indirect}{3},\citeproc{ref-Gonzalez2008}{23}{]}.

\subsection{Effect of Background Traffic}\label{sec-traffic-study}

The preceding analyses considered a single vehicle in isolation. In
practice, bridges carry multiple vehicles simultaneously, and the
resulting background traffic loads modify the bridge response amplitude.
This study investigates whether background traffic affects the relative
accuracy of the decoupled approach. The 30 m bridge (Eshkevari
configuration) is loaded with both the commercial vehicle and the heavy
truck as the instrumented sensing vehicle, while background traffic is
modelled as a uniformly distributed random nodal force whose total
magnitude corresponds to \(n\) vehicles of 2000 kg each, with
\(n = 0, 5, 10, 20\).

\begin{figure}[htbp]

\centering{

\includegraphics[width=0.75\linewidth,height=\textheight,keepaspectratio]{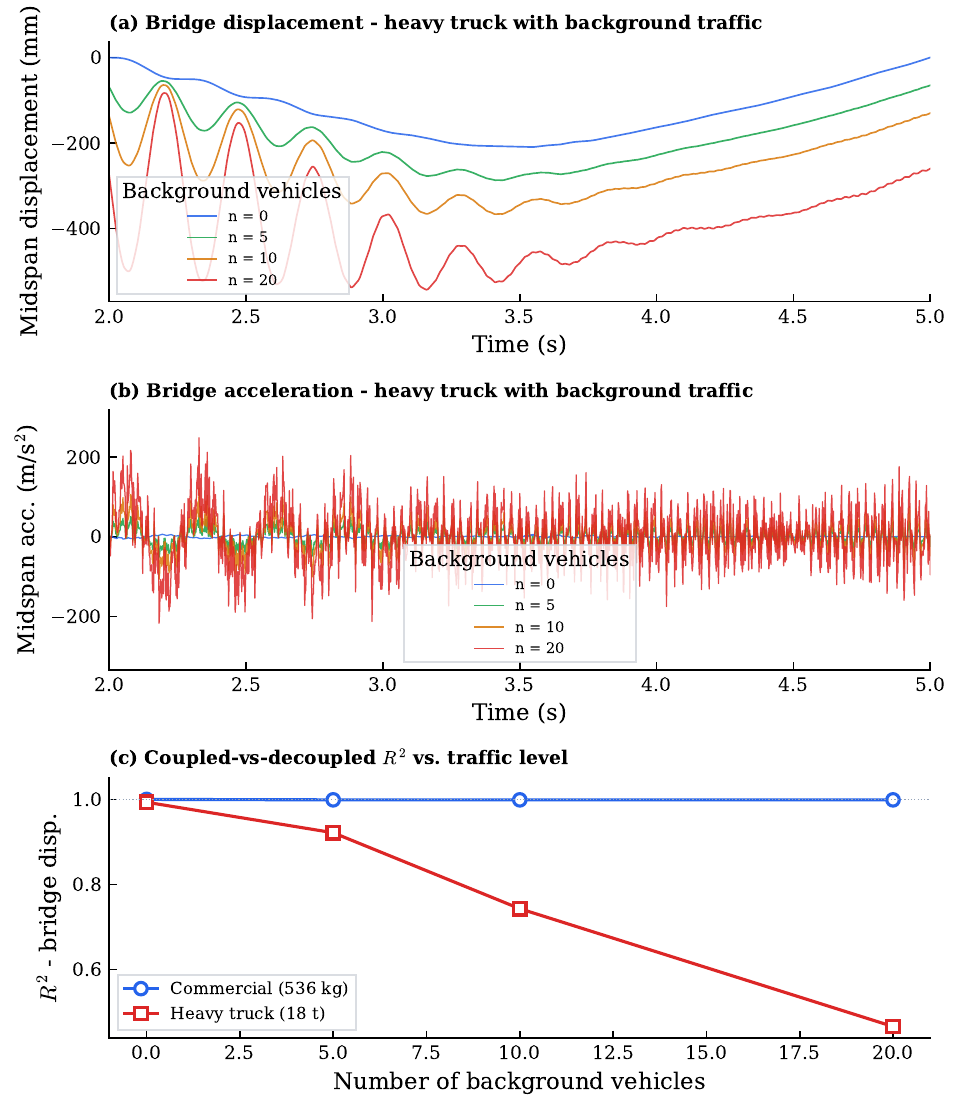}

}

\caption{\label{fig-traffic-study}Effect of background traffic on bridge
response (30 m bridge, heavy truck, coupled analysis). (a) Bridge
midspan displacement with increasing traffic levels. (b) Bridge midspan
acceleration with increasing traffic levels. (c) Coefficient of
determination $R^{2}$ between coupled and decoupled analyses as a
function of the number of background vehicles, for both vehicle types.}

\end{figure}%

Background traffic increases both the bridge displacement and
acceleration amplitudes substantially (Figure~\ref{fig-traffic-study} a,
b). The coefficient of determination $R^{2}$ between coupled and
decoupled analyses (Figure~\ref{fig-traffic-study} c) decreases with the
number of background vehicles for both vehicle types, dropping most
sharply from the isolated vehicle case (\(n = 0\)) to moderate traffic
(\(n = 5\)). For the commercial vehicle, $R^{2}$ stays close to unity
across all traffic levels, while for the heavy truck $R^{2}$ continues
to decrease, indicating progressively worse agreement as traffic
increases. The deterioration occurs because heavier total loading
produces larger bridge deflections, which in turn amplify the dynamic
contact forces that the decoupled approach neglects. Nevertheless, the
ordering between the two vehicle types is preserved at all traffic
levels: the heavy truck consistently produces lower $R^{2}$ than the
commercial vehicle, reinforcing the mass ratio as the primary
determinant of decoupled accuracy.

\subsection{Multi-Span Bridge
Configuration}\label{multi-span-bridge-configuration}

The framework supports multi-span continuous bridges by specifying
intermediate support locations. This capability is demonstrated using a
two-span bridge crossed by a fleet of ten vehicles. The fleet is
generated by perturbing a base two-axle composite half-car (comp2)
configuration: sprung mass, suspension stiffness, and suspension damping
are each independently sampled from a uniform distribution with
\(\pm 30\%\) range about their base values (\([0.7, 1.3]\times\) base).
Entry times are spaced along the approach segment so that vehicles enter
the bridge sequentially, and all vehicles travel at the same nominal
speed (20 m/s). The same random-fleet procedure is used in the
parametric studies that involve background traffic.

\begin{figure}[htbp]

\centering{

\includegraphics[width=0.75\linewidth,height=\textheight,keepaspectratio]{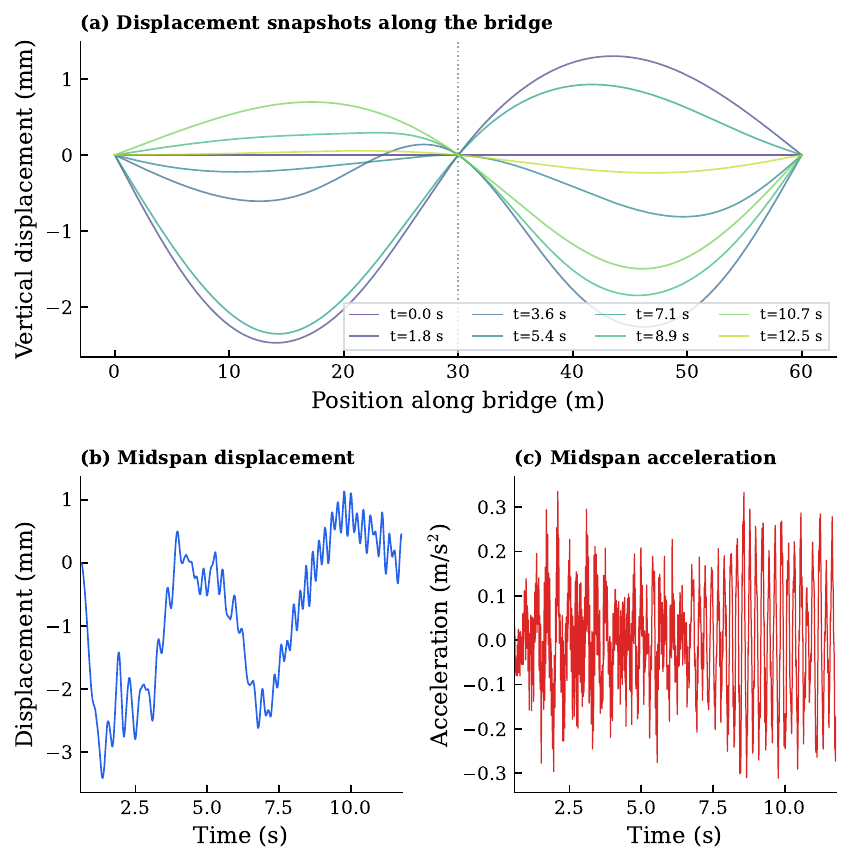}

}

\caption{\label{fig-multispan}Two-span continuous bridge (2 \(\times\)
30 m) under a fleet of 10 randomly generated vehicles: (a) bridge
displacement envelope over the full span showing the deflected shape at
different time instants; (b) midspan displacement time history of the
first span; (c) midspan acceleration time history of the first span.}

\end{figure}%

The two-span bridge analysis demonstrates the framework's capability to
handle continuous bridges with multiple vehicles. The displacement
snapshots show the characteristic shape of a continuous beam under
moving loads, with the inflection point at the intermediate support and
alternating deflections in the two spans as vehicles traverse them.

\section{Discussion}\label{sec-discussion}

The validation studies and parametric investigations presented in the
previous sections support several key observations about the proposed
framework and the VBI problem in general.

The framework's approach of using OpenSees as a black-box solver for
both the bridge and vehicle subsystems offers several practical
advantages. First, it enables researchers familiar with OpenSees to
perform VBI analyses without learning a new software package or
implementing custom finite element formulations. The bridge model can
incorporate any element type available in OpenSees, including nonlinear
beam-column elements for inelastic bridge response, fibre sections for
distributed plasticity, and specialised elements for soil-structure
interaction. Similarly, the vehicle model can be extended to include
nonlinear suspension behaviour or multi-body kinematics using the full
range of OpenSees materials and elements. Second, the rebuild-per-step
strategy, while computationally less efficient than maintaining a
persistent model, provides robustness: each time step starts from a
clean state, eliminating the risk of accumulated numerical errors or
state conflicts between the vehicle and bridge models.

The parametric study on the decoupled approach provides quantitative
guidance for practitioners. For the lightweight commercial vehicle (536
kg), the coefficient of determination $R^{2}$ between coupled and
decoupled bridge displacement remains above 0.999 across all four spans
tested (15--100 m), confirming that a vehicle of this mass has
negligible inertial effect regardless of bridge configuration. For the
heavy truck (18,000 kg), the agreement is noticeably weaker and drops
furthest on the 15 m bridge where the truck outweighs the bridge by a
factor of 24. The mass ratio \(\mu\) is the dominant predictor of
decoupled accuracy: the vehicle body acceleration $R^{2}$ follows the
same trend as the bridge displacement $R^{2}$ but with a larger gap
between vehicle classes, reflecting the amplification of interaction
effects through the suspension dynamics.

The speed study shows that higher vehicle speeds produce larger dynamic
amplification in both bridge and vehicle responses, consistent with the
well-known speed parameter effect in moving load problems. The roughness
study shows that surface condition substantially affects the vehicle
body response (Class E acceleration amplitudes exceed Class A by over an
order of magnitude) while the bridge midspan response remains
comparatively insensitive to roughness for this lightweight vehicle. The
traffic study reveals that background traffic weakens the
coupled-vs-decoupled agreement, with the heavy truck $R^{2}$ decreasing
noticeably as the number of background vehicles grows from 0 to 20,
while the commercial vehicle remains close to 1 throughout. The
deterioration occurs because the additional loading amplifies bridge
deflections and, consequently, the dynamic contact forces neglected by
the decoupled approach. Nevertheless, the ordering between vehicle types
is preserved at all traffic levels, reinforcing the mass ratio as the
primary determinant of decoupled accuracy
{[}\citeproc{ref-OBrien2015}{9},\citeproc{ref-Gonzalez2008}{23}{]}.

The open-source nature of the framework addresses a longstanding gap in
the VBI research community. While the theoretical foundations are well
established, the lack of shared, validated, and documented
implementations has led to unnecessary duplication of effort and made it
difficult to reproduce published results. By building on the widely used
OpenSees platform and providing a complete Python implementation with
benchmark configurations, the framework lowers the barrier to entry for
researchers and practitioners interested in VBI analysis and indirect
bridge monitoring.

\subsection{Limitations}\label{limitations}

The current implementation has several limitations that should be noted.
First, the framework is restricted to two-dimensional (planar) analysis,
neglecting lateral and torsional dynamics of both the bridge and the
vehicle. For straight, single-lane bridges loaded by vehicles travelling
along the centreline, this is a reasonable approximation; for curved
bridges, multi-lane loading, or vehicles with roll dynamics, a
three-dimensional extension would be necessary. Second, the bridge is
modelled with Euler-Bernoulli beam theory, which assumes that shear
deformation is negligible. For deep beams or short-span bridges,
Timoshenko beam elements would provide more accurate results. Third, the
tyre-road contact is modelled as a point contact with no loss of contact
(though wheel uplift is checked); a more detailed contact model would be
needed for rough roads at high speeds. Fourth, the rebuild-per-step
strategy incurs computational overhead that could be reduced by
maintaining persistent OpenSees models and updating only the
time-varying quantities. These limitations represent opportunities for
future development.

\section{Conclusions}\label{sec-conclusions}

This paper presented an open-source Python framework for vehicle-bridge
interaction analysis built on the OpenSees finite element platform. The
key contributions are:

\begin{enumerate}
\def\labelenumi{\arabic{enumi}.}
\item
  \textbf{A modular VBI framework using standard FE software.} The
  framework models the bridge and vehicle as independent OpenSees finite
  element subsystems coupled through an external iterative algorithm. No
  custom element formulations or source code modifications to OpenSees
  are required. This approach makes VBI analysis accessible to the broad
  community of OpenSees users and allows seamless integration with
  existing bridge modelling workflows.
\item
  \textbf{Support for multiple vehicle model types.} Five vehicle models
  of increasing complexity are implemented, from simple quarter-car
  spring-mass systems to half-car vehicles with body pitch, axle masses,
  and independent tyre-suspension assemblies. All models are constructed
  using standard OpenSees elements and driven by displacement excitation
  at the contact points.
\item
  \textbf{A validated iterative coupling algorithm.} The partitioned
  iterative scheme converges reliably in 1--4 iterations per time step
  with a tight tolerance of \(10^{-12}\) and reproduces benchmark
  results from the literature with high fidelity.
\item
  \textbf{A quantified simplified approach.} Systematic parametric
  studies across four bridge spans, two vehicle types spanning more than
  an order of magnitude in mass, six vehicle speeds, three roughness
  classes, and four traffic levels demonstrate that the
  vehicle-to-bridge mass ratio \(\mu\) is the dominant predictor of
  decoupled accuracy. For the lightweight commercial vehicle
  (\(\mu < 1\)), the coupled and decoupled bridge-displacement histories
  have $R^{2}$ above 0.999 across all spans. For heavy vehicles
  (\(\mu > 0.1\)), coupled analysis is necessary, with $R^{2}$ dropping
  noticeably below unity and reaching its lowest values when the
  vehicle mass approaches or exceeds the bridge mass.
\item
  \textbf{Open-source release.} The complete framework, including all
  vehicle configurations and benchmark scripts, is released as
  open-source software to promote reproducible research in
  vehicle-bridge dynamics and indirect bridge monitoring.
\end{enumerate}

Future work will extend the framework to three-dimensional analysis,
incorporate nonlinear material behaviour for both the bridge and vehicle
subsystems, and develop interfaces for real-time data assimilation in
drive-by bridge health monitoring applications.

\section*{Data Availability}\label{data-availability}
\addcontentsline{toc}{section}{Data Availability}

The complete source code and benchmark configurations presented in
this paper are openly available at
\url{https://github.com/MTalebi/OpenSees-VBI-Simulation}.

\section*{References}\label{references}
\addcontentsline{toc}{section}{References}

\phantomsection\label{refs}

\newpage{}

\section*{Appendix: OpenSees Vehicle Model Topology}\label{sec-appendix}
\addcontentsline{toc}{section}{Appendix: OpenSees Vehicle Model
Topology}

Table~\ref{tbl-vehicle-topology} summarises the OpenSees node layout,
element connectivity, and boundary conditions for each of the four
vehicle model types. In all models, axle nodes (contact points) are
fully fixed and excited through imposed displacement and velocity via
the \texttt{MultipleSupport} pattern. Non-axle nodes carry lumped mass
and are free to respond dynamically. Springs and dashpots are modelled
using \texttt{zeroLength} elements (1D models) or \texttt{twoNodeLink}
elements (2D models) with \texttt{Elastic} uniaxial materials whose
stiffness and damping arguments correspond to the spring constant \(k\)
and the viscous damping coefficient \(c\), respectively.

\begin{table}[H]
\centering
\caption{OpenSees node and element connectivity for each vehicle model
type. Axle nodes are constrained as fixed supports and driven by imposed
motion; non-axle nodes are free. Element notation \(i \to j\) denotes a
spring-dashpot connecting nodes \(i\) and \(j\).}
\label{tbl-vehicle-topology}
\small
\begin{tabular}{@{}
  >{\raggedright\arraybackslash}p{0.12\linewidth}
  >{\centering\arraybackslash}p{0.05\linewidth}
  >{\raggedright\arraybackslash}p{0.26\linewidth}
  >{\raggedright\arraybackslash}p{0.16\linewidth}
  >{\raggedright\arraybackslash}p{0.22\linewidth}
  >{\raggedright\arraybackslash}p{0.12\linewidth}@{}}
\toprule
Model & \texttt{ndm} & Nodes & Mass assignment & Elements & Notes \\
\midrule
\textbf{one\_axle\_simple} & 1 & 1 (axle, fixed), 2 (body) & \(m_v\) at
node 2 & \texttt{zeroLength} 1\(\to\)2 (\(k_v, c_v\)) & 2 DOFs \\
\textbf{one\_axle\_comp} & 1 & 1 (contact, fixed), 2 (axle), 3 (body) &
\(m_u\) at 2, \(m_s\) at 3 & \texttt{zeroLength} 1\(\to\)2 (\(k_t,
c_t\)); 2\(\to\)3 (\(k_s, c_s\)) & 3 DOFs \\
\textbf{two\_axle\_comp1} & 1 & 1,2 (axles, fixed); 3,4 (axle masses);
5,6 (body halves) & \(m_u\) at 3,4; \(m_s\) at 5,6 & Tyre: 1\(\to\)3,
2\(\to\)4; Susp: 3\(\to\)5, 4\(\to\)6 & 6 DOFs, independent halves \\
\textbf{two\_axle\_comp2} & 2 & 1,2 (axles, fixed); 5,6 (susp. tops);
7,8 (body at axle \(x\)); 9 (CG) & \(m_v, J_v\) at 9 & Susp:
1\(\to\)5, 2\(\to\)6 (\texttt{twoNodeLink}); Rigid links:
5\(\to\)7\(\to\)9, 6\(\to\)8\(\to\)9 (penalty \(\sim\!10^6 k_s\)) & 8
DOFs (2 measured: bounce + pitch) \\
\textbf{two\_axle\_comp3} & 2 & 1,2 (contacts, fixed); 3,4 (axle
masses); 5,6 (susp. tops); 7,8 (body at axle \(x\)); 9 (CG) &
\(m_{u,r}\) at 3, \(m_{u,f}\) at 4, \(m_v, J_v\) at 9 & Tyre:
1\(\to\)3, 2\(\to\)4; Susp: 3\(\to\)5, 4\(\to\)6; Rigid links:
5\(\to\)7\(\to\)9, 6\(\to\)8\(\to\)9 & 10 DOFs (3 measured: 2 axles +
bounce + pitch) \\
\bottomrule
\end{tabular}
\end{table}

For the half-car models with pitch (comp2 and comp3), the rigid-body
connection between the CG node (node 9) and the suspension tops (nodes
5, 6) is implemented using intermediate nodes (7, 8) located at the axle
\(x\)-coordinates but at body height. The \texttt{twoNodeLink} elements
connecting nodes 7→9 and 9→8 carry penalty stiffness in all three DOFs
(\(x\)-translation, \(y\)-translation, and rotation), ensuring that the
body moves as a rigid bar between the two suspension attachment points.
The \texttt{zeroLength} elements connecting nodes 5→7 and 6→8 similarly
enforce kinematic compatibility between the suspension tops and the body
endpoints. This penalty-based approach avoids the need for multi-point
constraints and is compatible with OpenSees' standard analysis
procedures.

\subsection*{Framework Module
Pseudocodes}\label{framework-module-pseudocodes}
\addcontentsline{toc}{subsection}{Framework Module Pseudocodes}

The following pseudocodes summarise the supporting framework modules as
step-by-step procedures for readers who prefer the algorithmic form.
They restate, in algorithmic notation, the operations described
narratively in Sections~\ref{bridge-finite-element-model},
\ref{vehicle-finite-element-models}, \ref{force-mapping}, and
\ref{road-roughness-generation}. The main coupled and decoupled
procedures appear as Algorithms~\ref{alg:iterative} and
\ref{alg:decoupled} in Section~\ref{sec-methodology}.

\begin{algorithm}[H]
\caption{Bridge Finite Element Construction and Time Stepping}
\label{alg:bridge-fe}
\begin{algorithmic}[1]
\Require Span $L$, support locations, $EI$, $\bar{m}$, $\zeta$, discretisation $N_e$, initial conditions $\{w_n, \dot{w}_n, \ddot{w}_n\}$, nodal forces $\mathbf{F}$, time step $\Delta t$
\Ensure Bridge response at $t_{n+1}$: $\{w_{n+1}, \dot{w}_{n+1}, \ddot{w}_{n+1}\}$
\State \texttt{ops.wipe()} \Comment{Clear previous model}
\State Define $N_n = N_e + 1$ nodes at spacing $\Delta x = L / N_e$
\State Apply boundary conditions: pin at first support, roller(s) at remaining
\State Assign lumped mass: $m_j = \bar{m}\,\Delta x$ (interior nodes), $\bar{m}\,\Delta x/2$ (end nodes)
\State Define \texttt{elasticBeamColumn} elements with $A$, $E$, $I$
\State Eigenvalue analysis: $\omega_1, \omega_2 \gets$ \texttt{ops.eigen}$(N_{\text{modes}})$
\State Rayleigh damping: $\alpha_M, \beta_K \gets$ solve $\zeta = \tfrac{1}{2}(\alpha_M/\omega_i + \beta_K \omega_i)$ for $i = 1, 2$
\State \texttt{ops.rayleigh}$(\alpha_M, \beta_K, 0, 0)$
\State Apply nodal forces $\mathbf{F}$ via \texttt{Path} time series and \texttt{Plain} load pattern
\State Set initial conditions: \texttt{setNodeDisp}, \texttt{setNodeVel}, \texttt{setNodeAccel} at each node
\State Configure \texttt{Newmark} integrator ($\gamma = 0.5$, $\beta = 0.25$)
\State \texttt{ops.analyze}$(1, \Delta t)$
\State Extract updated $\{w_{n+1}, \dot{w}_{n+1}, \ddot{w}_{n+1}\}$ via \texttt{nodeDisp}, \texttt{nodeVel}, \texttt{nodeAccel}
\end{algorithmic}
\end{algorithm}

\begin{algorithm}[H]
\caption{Vehicle Finite Element Construction and Time Stepping}
\label{alg:vehicle-fe}
\begin{algorithmic}[1]
\Require Vehicle properties (masses, stiffnesses, dampings), model type, initial conditions, axle displacements $\{u_i\}$ and velocities $\{\dot{u}_i\}$
\Ensure Vehicle response at $t_{n+1}$, axle reaction forces $\{F_{a,i}\}$
\State \texttt{ops.wipe()}
\If{model $\in$ \{one\_axle\_simple, one\_axle\_comp\}}
    \State Define 1D model (\texttt{ndm}$= 1$)
    \State Create axle contact node(s), fix all DOFs
    \State Create body/axle-mass nodes with lumped masses
    \State Define \texttt{zeroLength} spring-dashpot elements ($k, c$)
\ElsIf{model $\in$ \{two\_axle\_comp2, two\_axle\_comp3\}}
    \State Define 2D model (\texttt{ndm}$= 2$, \texttt{ndf}$= 3$)
    \State Create axle contact nodes, fix all DOFs
    \State Create CG node with body mass $m_v$ and rotational inertia $J_v$
    \State Create intermediate nodes at axle $x$-coordinates
    \State Define \texttt{twoNodeLink} suspension elements
    \State Define penalty links ($k_{\text{pen}} \approx 10^6 k_s$) for rigid-body connection
\EndIf
\State Apply \texttt{MultipleSupport} with \texttt{imposedMotion} at each axle node
\State Prescribe displacement $u_i(t)$ and velocity $\dot{u}_i(t)$ via \texttt{Path} time series
\State Set initial conditions from previous time step
\State Configure \texttt{Newmark} integrator ($\gamma = 0.5$, $\beta = 0.25$)
\State \texttt{ops.analyze}$(1, \Delta t)$
\State \texttt{ops.reactions}(\texttt{-dynamic}, \texttt{-rayleigh})
\State $F_{a,i} \gets$ \texttt{nodeReaction}$(i)$ for each axle node
\end{algorithmic}
\end{algorithm}

\begin{algorithm}[H]
\caption{ISO 8608 Road Roughness Profile Generation}
\label{alg:roughness}
\begin{algorithmic}[1]
\Require Roughness class coefficient $G_d(n_0)$, spatial range $[x_0, x_{\max}]$, mesh size $\Delta x$, bridge span $L$
\Ensure Roughness profile $r(x_j)$
\State $\Delta n \gets \min(0.01, \; 1/(2L))$ \Comment{Avoid periodicity within bridge span}
\State $n_l \gets \min(0.01, \; \Delta n)$, \quad $n_u \gets 10$ cycles/m
\State Generate spatial frequency array: $n_k = n_l, \; n_l + \Delta n, \; \ldots, \; n_u$
\State Compute PSD: $G_d(n_k) = G_d(n_0) \cdot (n_k / n_0)^{-2}$, \quad $n_0 = 0.1$ cycles/m
\State Compute amplitudes: $A_k = \sqrt{2 \, G_d(n_k) \, \Delta n}$
\State Generate random phases: $\phi_k \sim \mathcal{U}[0, 2\pi]$
\State Generate position array: $x_j = x_0, \; x_0 + \Delta x, \; \ldots, \; x_{\max}$
\For{each position $x_j$}
    \State $r(x_j) = \sum_k A_k \cos(2\pi \, n_k \, x_j + \phi_k)$
\EndFor
\State \textbf{Optional:} Apply moving-average smoothing (window size $w$)
\end{algorithmic}
\end{algorithm}

\begin{algorithm}[H]
\caption{Vehicle-to-Bridge Force Mapping}
\label{alg:force-map}
\begin{algorithmic}[1]
\Require Axle positions $\{x_{a,i}\}$, dynamic reaction forces $\{F_{a,i}\}$, static axle weights $\{W_i\}$, bridge node coordinates, bridge span $L$
\Ensure Equivalent nodal force vector $\mathbf{F}_{\text{inter}}$
\State Initialise $\mathbf{F}_{\text{inter}} \gets \mathbf{0}$
\For{each axle $i$}
    \State $F_{\text{total},i} \gets W_i + F_{a,i}$ \Comment{Static weight + dynamic reaction}
    \If{$x_{a,i} < 0$ \textbf{or} $x_{a,i} > L$} \Comment{Axle off bridge}
        \State \textbf{continue}
    \EndIf
    \If{$F_{\text{total},i} > 0$} \Comment{Wheel uplift detected}
        \State Set $F_{\text{total},i} \gets 0$ and flag warning
    \EndIf
    \State Find element $[x_j, \, x_{j+1}]$ containing $x_{a,i}$
    \State $\xi \gets (x_{a,i} - x_j) / (x_{j+1} - x_j)$ \Comment{Local coordinate}
    \State $F_j \gets F_j + (1 - \xi) \, F_{\text{total},i}$
    \State $F_{j+1} \gets F_{j+1} + \xi \, F_{\text{total},i}$
\EndFor
\end{algorithmic}
\end{algorithm}

\end{document}